\pdfoutput=1

\documentclass[10pt,twocolumn,twoside,journal]{IEEEtran}

\usepackage{cs}
\usepackage{mathsymb}
\usepackage{verbatim}
\usepackage{subfigure}

\usepackage{graphicx}
\graphicspath{{./}{./Fig/}{./Figs/}}

\usepackage{algorithm2e}
\let\savedalgorithm\algorithm
\let\savedendalgorithm\endalgorithm
\newenvironment{algorithmic}{%
\savedalgorithm
}{%
\savedendalgorithm
}

\usepackage{algorithm}

\usepackage{amsmath}
\usepackage{url}
\usepackage{multirow}

\def\Integer{\mathbb{Z}^+}

\begin{document}

\title{Efficiently Learning a Detection Cascade \\
       with Sparse Eigenvectors}

\author{ 
         Chunhua~Shen,
         Sakrapee Paisitkriangkrai,
         and
         Jian Zhang,~\IEEEmembership{Senior Member,~IEEE}
\thanks
{ 
NICTA is funded through the Australian Government's 
{\em Backing Australia's Ability} initiative, 
in part through the Australian Research Council.           
The associate editor coordinating the review of this manuscript
and approving it for publication was Dr. X X. 
}
\thanks
{
C. Shen is with NICTA, Canberra Research Laboratory, 
Locked Bag 8001, Canberra, ACT 2601, Australia,
and also with the Australian National University, Canberra,
ACT 0200,Australia
(e-mail: chunhua.shen@nicta.com.au).
}
\thanks
{
S. Paisitkriangkrai
and J. Zhang are with NICTA, Neville Roach Laboratory,
Kensington, NSW 2052, Australia, and also with the University of New
South Wales, Sydney, NSW 2052, Australia
(e-mail: \{paul.pais, jian.zhang\}@nicta.com.au). 
}
}


\maketitle

\begin{abstract}

      Real-time object detection has many applications in video
      surveillance, teleconference and
      multimedia retrieval \etc. Since Viola and Jones
      \cite{Viola2004Robust} proposed the first real-time AdaBoost
      based face detection system, much effort has been spent on
      improving the boosting method.
      In this work, we first show that
      feature selection methods other than boosting can also be used
      for training an efficient object detector.  In particular, we
      introduce Greedy Sparse Linear Discriminant Analysis (GSLDA)
      \cite{Moghaddam2007Fast} for its conceptual simplicity and 
      computational efficiency; and
      slightly better detection performance is achieved compared with
      \cite{Viola2004Robust}. 
      Moreover, we propose a new technique,
      termed Boosted Greedy Sparse Linear Discriminant Analysis
      (BGSLDA), to efficiently train a detection cascade.
      BGSLDA exploits
      the sample re-weighting property of boosting and the
      class-separability criterion of GSLDA. Experiments
      in the domain of highly skewed data distributions, \eg, face
      detection, demonstrates that classifiers trained with the
      proposed BGSLDA outperforms AdaBoost and its variants.  This finding
      provides a significant opportunity to argue that AdaBoost and
      similar approaches are not the only methods that can achieve
      high classification results for high dimensional data in 
      object detection.

\end{abstract}

\begin{IEEEkeywords}
        Object detection, 
        AdaBoost,
        asymmetry, 
        greedy sparse linear discriminant analysis,
        feature selection, 
        cascade classifier.
\end{IEEEkeywords}

\section{Introduction}

        \IEEEPARstart{R}{eal-time} objection detection such as face
        detection has numerous computer vision applications, \eg, 
        intelligent video surveillance, vision based teleconference
        systems and content based image retrieval.
        Various detectors have been proposed in the literature
        \cite{Viola2004Robust,Rowley1998NN,Romdhani2001Detection}.
        Object detection is
        challenging due to the variations of the visual appearances,
        poses and illumination conditions. 
        Furthermore, object detection is a
        {\em highly-imbalanced} 
        classification task. 
        A typical natural image contains many more negative background
        patterns than object patterns.  The number of background
        patterns can be $100,000$ times larger than the number
        of object patterns. That means, if one wants to achieve a high
        detection rate,
        together with a low false detection rate, one
        needs a specific classifier. The cascade classifier takes this
        imbalanced distribution into consideration
        \cite{Viola2002Fast}.   
        Because of the huge success of Viola and Jones' real-time
        AdaBoost based face detector \cite{Viola2004Robust}, a lot of
        incremental work has been proposed. Most of them have focused
        on improving the underlying boosting method or accelerating
        the training process. 
        For example, AsymBoost was introduced in \cite{Viola2002Fast}
        to alleviate the limitation of AdaBoost in the context of
        highly skewed example distribution.  Li \etal
        \cite{Li2000Performance} proposed FloatBoost for a better
        detection accuracy by introducing a backward feature
        elimination step into the AdaBoost training procedure.  Wu
        \etal \cite{Wu2008Fast} used forward feature selection for
        fast training by ignoring the re-weighting scheme in AdaBoost.
        Another technique based on the statistics of the weighted
        input data was used in \cite{Pham2007Fast} for even faster 
        training.
         KLBoost was proposed in \cite{Liu2003KL} to train a strong
        classifier. The weak classifiers of KLBoost are based on
        histogram divergence of linear features. Therefore in the
        detection phase, it is not as efficient as Haar-like features.
        Notice that in KLBoost, the classifier design is separated
        from feature selection. 
        In this work (part of which was published
        in preliminary form in \cite{Paisitkriangkrai2009CVPR}), 
        we propose an improved learning
        algorithm for face detection, dubbed Boosted Greedy Sparse
        Linear Discriminant Analysis (BGSLDA).

\begin{figure}[t]
            \begin{center}
                \includegraphics[width=0.48\textwidth]{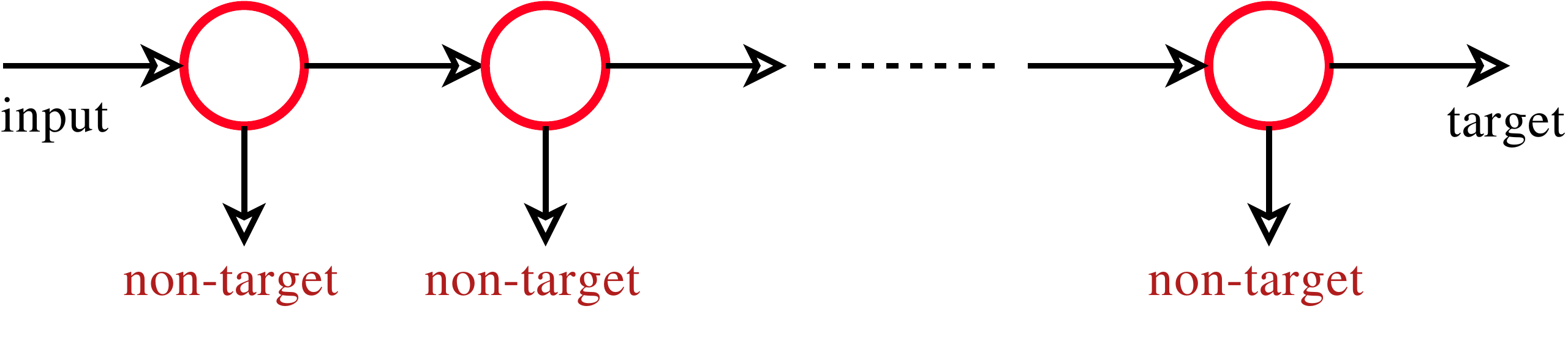}
            \end{center}
            \caption{A cascade classifier with multiple  nodes. 
            Here a circle represents
            a node classifier. An input patch is classified as a target only when
            it passes tests at each node classifier.}
            \label{fig:cas}
        \end{figure}

        Viola and Jones \cite{Viola2004Robust} introduced a framework
        for selecting discriminative features and training classifiers
        in a cascaded manner as shown in Fig.~\ref{fig:cas}.
        The cascade framework allows most
        non-face patches to be rejected quickly before reaching the
        final node, resulting in fast performance. 
        A test image patch is reported as a face
        only if it passes tests in all nodes. This way,
        most non-face patches
        are rejected by these early nodes.
        Cascade detectors lead to very fast
        detection speed and high detection rates.
        Cascade classifiers have also been used in the context of
        support vector machines (SVMs) for faster face detection
        \cite{Romdhani2001Detection}. 
        In \cite{Bourdev05SoftCascade}, soft-cascade is developed to 
        reduce the training and design complexity. 
        The idea was further developed in \cite{Pham2008Detection}. 
        We have followed Viola and Jones' original cascade classifiers
        in this work.

        One issue that contributes to the efficacy of
        the system comes from the use of AdaBoost algorithm for training 
        cascade nodes. AdaBoost is a forward stage-wise additive
        modeling with the weighted exponential loss function. The
        algorithm combines an ensemble of weak classifiers to produce
        a final strong classifier with high classification accuracy. 
        AdaBoost chooses a small subset of weak classifiers and assign them
        with proper coefficients. The linear combination of weak classifiers
        can be interpreted as a decision hyper-plane in the weak classifier
        space. 
        The proposed BGSLDA differs from the original
        AdaBoost in the following aspects. Instead of selecting decision stumps
        with minimal weighted error as in AdaBoost, the proposed algorithm
        finds a new weak leaner that maximizes the class-separability
        criterion. As a result, the coefficients of selected weak classifiers
        are updated repetitively during the learning process according to this
        criterion.
%

        Our technique  differs from \cite{Wu2008Fast} in the following 
        aspects. \cite{Wu2008Fast} proposed the concept of Linear Asymmetric
        Classifier (LAC) by addressing the asymmetries and 
        {\em asymmetric
         node learning} goal in the cascade framework. Unlike our work
        where the features are selected based on the 
        Linear Discriminant Analysis (LDA) criterion,
        \cite{Wu2008Fast} selects features using AdaBoost$/$AsymBoost
        algorithm. Given the selected features, Wu \etal then build
        an optimal
        linear classifier for the node learning goal
        using LAC or LDA. 
        Note that similar techniques have
         also been applied in neural network. In
         \cite{Webb1990Optimised},
         a nonlinear adaptive feed-forward layered network with
         linear output units has been introduced.
         The input data is nonlinearly
         transformed into a space in which classes can be separated
         more easily. 
        Since LDA considers the number of training
        samples of each class, applying LDA at the output of neural
        network hidden units has been shown to increase the
        classification accuracy of two-class problem with unequal
        class membership.
        As our experiments show, in terms of feature selection, 
        the proposed BGSLDA methods is superior than AdaBoost and
        AsymBoost  
        for object detection.


The key contributions of this work are as follows. 
\begin{itemize}
    \item
           We introduce GSLDA  as an alternative approach for training
           face detectors. Similar results are obtained compared with
           Viola and Jones' approach.
\item
            We propose a new algorithm, BGSLDA, which combines the
            sample re-weighting schemes typically used in boosting
            into GSLDA. Experiments show that BGSLDA can achieve
            better detection performances. 
\item
            We show that feature selection and classifier training
            techniques can have different objective functions (in
            other words, the two processes can be separated) in the
            context of training a visual detector. This offers more
            flexibility and even better performance.  Previous
            boosting based approaches select features and train a
            classifier simultaneously. 
\item
            Our results confirm that it is beneficial to consider the
            highly skewed data distribution when training a detector.
            LDA's learning criterion already incorporates this
            imbalanced data information. Hence it is better than
            standard AdaBoost's exponential loss for training an
            object detector. 
\end{itemize}


        The remaining parts of the paper are structured as follows. In
        Section~\ref{sec:gslda}, the GSLDA algorithm is introduced as
        an alternative learning technique to object detection
        problems. We then discuss how LDA incorporates imbalanced
        data information when training a classifier in
        Section~\ref{sec:asym}.  Then, in Sections~\ref{sec:bgslda}
        and \ref{sec:complexity}, the proposed BGSLDA algorithm is
        described and the training time complexity is discussed.
        Experimental results are shown in Section~\ref{sec:exp} and
        the paper is concluded in Section~\ref{sec:conclusion}.

\section{Algorithms}
\label{sec:algo}

        In this section, we present alternative techniques to AdaBoost
        for object detection.  We start with a short explanation of
        the concept of GSLDA \cite{Moghaddam2006Generalized}.  Next,
        we show that like AsymBoost \cite{Viola2002Fast},
        LDA is better at handling asymmetric data than AdaBoost. We
        also propose the new algorithm that makes use of sample
        re-weighting scheme commonly used in AdaBoost to select a
        subset of relevant features for training the GSLDA classifier.
        Finally, we analyze the training time complexity of the
        proposed method.

\subsection{Greedy Sparse Linear Discriminant Analysis}
\label{sec:gslda}

        Linear Discriminant Analysis (LDA) can be cast as a
        generalized eigenvalue decomposition. Given a pair of
        symmetric matrices corresponding to the between-class ($S_b$)
        and within-class covariance matrices ($S_w$), one maximizes a
        class-separability criterion defined by the generalized
        Rayleigh quotient: 
\begin {equation}
\label{EQ:GRQ}
   \max_{\bw} \,\,  \frac{ \bw^\T S_b \bw }{ \bw^\T S_w \bw }.
\end{equation}
        The optimal solution of a generalized Rayleigh quotient is the
        eigenvector corresponding to the maximal eigenvalue. 
        The sparse version of LDA is to solve \eqref{EQ:GRQ}
        with an additional sparsity constraint:
        \begin{equation}
            \label{EQ:SPConstraint}
            \card(\bw) = k,
        \end{equation}
        where $ \card(\cdot)$ counts the number of nonzero components, \aka
        the $ \ell_0$ norm. $ k \in \Integer $ is an integer set by a user. 
        Due to this sparsity constraint, the problem
        becomes non-convex and NP-hard.
        In
        \cite{Moghaddam2006Generalized}, Moghaddam \etal presented a
        technique to compute optimal sparse linear discriminants using
        branch and bound approach. Nevertheless, finding the exact
        global optimal solutions for high dimensional data is
        infeasible. The algorithm was
        extended in \cite{Moghaddam2007Fast},
        with new sparsity bounds and efficient
        matrix inverse techniques to speed up the computation time by
        $1000\times$. The technique works by
        sequentially adding the new variable which yields the maximum
        eigenvalue (forward selection) until the maximum number of
        elements are selected or some predefined condition is met. As
        shown in \cite{Moghaddam2007Fast}, for two-class
        problem, the computation can be made very efficient as
        the only finite eigenvalue $ \lambda_{\max} ( S_b, S_w    )   $
        can be computed in closed-form as 
        $  {\boldsymbol b}^\T S_w ^{-1} {\boldsymbol b}   $ with
        $ S_b = {\boldsymbol b} {\boldsymbol b} ^\T  $ because in this case
        $ S_b $ is a rank-one matrix. $ {\boldsymbol b}$ is a column vector.
        Therefore, the computation is mainly determined by the
        inverse of $ S_w$. When a greedy approach is adopted to sequentially
        find the suboptimal $ \bw $,
        a simple rank-one update for computing $ S_w^{-1}$ significantly 
        reduces the computation complexity \cite{Moghaddam2007Fast}.
        We have mainly used forward greedy search in this work.
        For forward greedy search, 
        if $ l $ is the current subset of
        $ k $ indices and $ m = l \cup   i $ for  candidate $ i $ which is not
        in $ l $. 
        The new augmented inverse $ ( S_w^m )^{-1}   $ can be calculated 
        in a fast way by recycling the last step's result
        $ ( S_w^l )^{-1}   $:
        \begin{equation}
            \label{EQ:rank1}
            ( S_w^m ) ^ { -1 }  = 
             \begin{bmatrix}
                                (S_w^l)^{-1}   +  a_i \bu_i \bu_i ^\T  &  - a_i \bu_i \\
                                - a_i \bu_i                          &  a_i
             \end{bmatrix},
        \end{equation}
        where $ \bu_i = ( S_w^l )^{-1}  S_{w,li}  $ with $ (li) $ indexing the $ l $ rows and 
        $ i$-th column of $ S_w $ and $ a_i = 1 /  (  S_{w,ii} - S_{w,li} ^ \T \bu_i )$
        \cite{Golub1996Matrix,Moghaddam2007Fast}.
        
%
%
        Note that we have experimented with other sparse 
        linear regression and classification
        algorithms, \eg, $\ell_1$-norm linear support vector machines,
        $\ell_1$-norm regularized
        log-linear models, \etc.
        However, the major drawback of these techniques is that they do not
        have an {\em explicit} parameter that controls the number of features to be selected. 
        The trade-off parameter (regularization parameter) only controls the degree of sparseness.
        One has to tune this parameter using cross-validation.  
        Also $\ell_1$ penalty methods often lead to sub-optimal 
        sparsity \cite{Zhang08Multistage}.
        Hence, we have decided to apply GSLDA,
        which makes use of greedy feature selection and the number of features can be predefined.
        It would be of interest to compare our method with $ \ell_1 $-norm induced 
        sparse models \cite{Destrero2009Sparsity}.

        The following paragraph explains how we apply GSLDA classifier
        \cite{Moghaddam2007Fast} as an alternative feature selection
        method to classical Viola and Jones' framework
        \cite{Viola2004Robust}.

        Due to space limit, we omit the explanation of cascade classifiers.
        Interested readers should refer to \cite{Viola2004Robust,
        Wu2008Fast} for details.
        The GSLDA object detector operates as follows. The set of
        selected features is initialized to an empty set. The first
        step (lines $4-5$) is to train
        weak classifiers, for example, 
        decision stumps on Haar features.\footnote{We
        introduce nonlinearity into our system by applying decision
        stump learning to raw Haar feature values. By nonlinearly
        transforming the data, the input can now be separated more
        easily using simple linear classifiers. 
        
        Note that any
        classifiers can be applied here. We also use LDA on covariance features
        for human detection as described \cite{Paisitkriangkrai2008Fast}.
        For the time being, we focus on decision stumps on Haar-like features. 
        We will give details about covariance features later.}
        For each Haar-like rectangle feature, the threshold
        that gives the minimal classification error is stored 
        into the lookup
        table. In order to achieve maximum class separation, the
        output of each decision stump is examined and the decision stump
        whose output yields the maximum eigenvalue is sequentially
        added to the list (line $7$, step ($1$)).  The process
        continues until the predefined condition is met (line $6$).

        The proposed GSLDA based detection framework is summarized in
        Algorithm~\ref{ALG:GSLDA}.

%
%
%
%
\def\dmin{ D_{\mathrm{min}} }
\def\fmax{ F_{\mathrm{max}} }
\def\ftarget{ F_{\mathrm{target}} }

\SetVline
\linesnumbered

\begin{algorithm}[t]
\caption{The training procedure for building a cascade of GSLDA object detector.}
\begin{algorithmic}
\small{
   \KwIn{
   \begin{itemize}
      \item
         A positive training set and a negative training set;
      \item
         A set of Haar-like rectangle features $h_1, h_2, \cdots$;
      \item
         $\dmin$:  minimum acceptable detection rate per cascade level;
      \item
         $\fmax$:  maximum acceptable false positive rate per cascade
         level;
      \item
         $ \ftarget $: target overall false positive rate;
   \end{itemize}
   }
   { {\bf Initialize}:
	$i = 0$;  $D_i = 1$;  $ F_i = 1$;}

\While{ $  \ftarget < F_i $ }
{
$ i = i + 1$; $ f_i = 1 $;
   
   \ForEach{feature}
   {
     Train a weak classifier (\eg,
     a decision stump parameterized by a threshold $ \theta $) 
     with the 
     smallest error on the training set;
   }

   \While{$ f_i > \fmax$ }
      { 
      
         1. Add the best weak classifier (\eg, decision stump)
         that yields
         the maximum class separation;
         \\
         2. Lower classifier threshold such that $\dmin$ holds;
         \\
         3. Update $f_i$ using this classifier threshold;
      }
      $ D_{i+1} = D_i \times \dmin $; $ F_{i+1} = F_i \times f_i $;
      and remove correctly classified negative samples from the training
      set;
      
      \If{ $\ftarget < F_i $ }
      {Evaluate the current cascaded classifier on the negative images
      and add misclassified samples into the negative training set;}
}
\KwOut{
\begin{itemize}
   \item
      A cascade of classifiers for each cascade
      level $i = 1,\cdots$; 
   \item
      Final training accuracy: $F_i$ and $D_i$;
\end{itemize}
}
%
%
}
\end{algorithmic}
\label{ALG:GSLDA}
\end{algorithm}

\subsection{Linear Discriminant Analysis on Asymmetric Data}
\label{sec:asym}

        In the cascade classifiers, we would prefer to have a
        classifier that yields high detection rates without
        introducing many false positives.  Binary variables (decision
        stump outputs) take the Bernoulli distribution and it can be
        easily shown that the log likelihood ratio is a linear
        function. In the Bayes sense, linear classifiers are optimum
        for normal distributions with equal covariance matrices.
        However, due to its simplicity and robustness, linear
        classifier has shown to perform well not only for normal
        distributions with unequal covariance matrices but also
        non-normal distributions. A linear classifier can be written
        as
   \begin{equation}
             \label{EQ:linear}
                F( \bx ) =
                \begin{cases}
                +1 & \textrm{if  } \sum_{t=1}^n {w_t h_t(\bx)} + \theta \geq 0;\\
                -1 & \textrm{otherwise},
                \end{cases}
   \end{equation}
where $h(\cdot)$ defines a function which returns binary outcome, $
\bx $ is the input image features and $\theta$ is an optimal threshold
such that the minimum number of examples are misclassified. In this
paper, our linear classifier is the summation of decision stump
classifiers. By central limit theorem, the linear classifier is close
to normal distribution for large $n$.

The asymmetric goal for training cascade classifiers can be written as
a trade-off between false acceptance rate $\varepsilon_1$ and false
rejection rate $\varepsilon_2$ as
\begin{equation}
  \label{EQ:asymgoal}
  r = \varepsilon_1 + \mu \varepsilon_2,
\end{equation}
where $\mu$ is a trade-off parameter.  
The objective of LDA is to maximize the projected between-class
covariance matrix (distance between the mean of two classes) and
minimize the within-class covariance matrix (total covariance matrix).
The selected weak classifier is guaranteed to achieve this goal.
Having large projected mean difference and small projected class
variance indicates that the data can be separated more easily and,
hence, the asymmetric goal can also be achieve more easily. On the
other hand, AdaBoost minimizes symmetric exponential loss function
that does not guarantee high detection rates with little false
positives \cite{Viola2002Fast}. The selected features are therefore no
longer optimal for the task of rejecting negative samples.

Another way to think of this is that AdaBoost sets initial positive
and negative sample weights to $0.5/N_p$ and $0.5/N_n$ ($N_p$ and
$N_n$ is the number of positive samples and negative samples). The
prior information about the number of samples in each class is then 
{\em completely lost} during training.
In contrast, LDA takes the number of samples in each class into
consideration when solving the optimization problem, \ie, the number of
samples is used in calculating the between-class covariance matrix
($S_B$). Hence, $S_B$ is the weighted difference between class mean
and sample mean. 
\begin{equation}
  \label{EQ:S_B}
  S_B = \sum_{c_i} N_{c_i}(\mu_{c_i} - \overline{x})(\mu_{c_i} -
  \overline{x})^\T,
\end{equation}
where
        $\mu_{c_i} = N_{c_i}^{-1} \sum_{j \in c_i} x_j$; $\overline{x}
        = N^{-1} \sum_j x_j$; $N_{c_i}$ is the number of samples in
        class $c_i$ and $N$ is the total number of samples. This extra
        information minimizes the effect of imbalanced data set.

\begin{figure}
	  \begin{center}
        \subfigure[]
        {
		        \includegraphics[width=0.4\textwidth,clip]{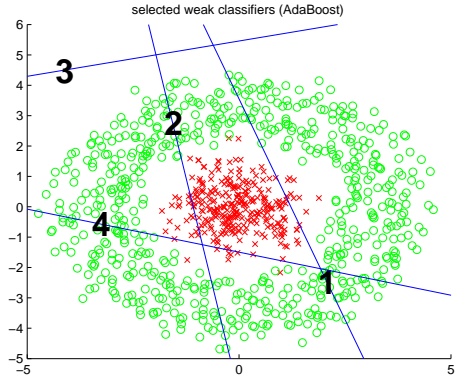}
        }
		\subfigure[]
        {
		        \includegraphics[width=0.4\textwidth,clip]{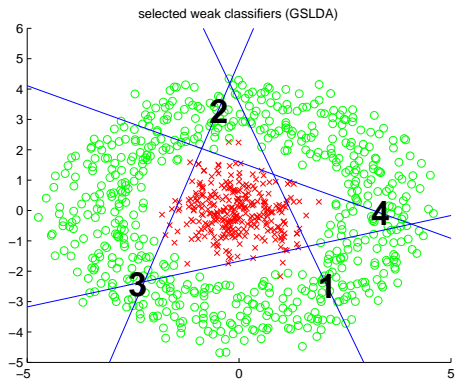}
        }
		\end{center}
		\caption{
		Two examples on a toy data set:
        (a) AdaBoost classifier; (b) GSLDA classifier (forward pass).
		$\times$'s and $ \circ $'s represent positive and negative
        samples, respectively. 
		Weak classifiers are plotted as lines. The number on the line indicates
		the order in which weak classifiers are selected. 
        AdaBoost
        selects weak classifiers for attempting to balance weighted
        positive and negative error.
        Notice that 
		AdaBoost's third weak classifier classifies all samples as
		negative due to the very small positive sample weights.
		In contrast, GSLDA selects weak classifiers based on the maximum
		class separation criterion. We see that 
        four weak classifiers of GSLDA model the positives well and most
        of the negative are rejected.
		}
	 \label{fig:asymdata}
  \end{figure}

        In order to demonstrate this, we generate an artificial data
        set similar to one used in \cite{Viola2002Fast}. We learn a
        classifier consisting of $4$ linear classifiers and the
        results are shown in Fig.~\ref{fig:asymdata}. From the
        figure, we see that the first weak classifier (\#1) selected by both
        algorithms are the same since it is the only linear classifier
        with minimal error. AdaBoost then re-weights the samples and
        selects the next classifier (\#2) which has the smallest
        weighted error. From the figure, the second weak classifier
        (\#2) introduces more false positives to the final classifier.
        Since most positive samples are correctly classified, the
        positive samples' weights are close to zero. AdaBoost selects
        the next classifier (\#3) which classifies all samples as
        negative.  
        Therefore it is clear that all but the first weak classifier
        learned by AdaBoost are poor because it tries to balance
        positive and negative errors. The final combination of these
        classifiers are not able to produce high detection rates without
        introducing many false positives.
        In contrast to AdaBoost, GSLDA selects the second
        and third weak classifier (\#2, \#3) based on the maximum
        class separation criterion. Only the linear classifier whose
        outputs yields the maximum distance between two classes is
        selected.  As a result, the selected linear classifiers
        introduce much less false positives
        (Fig.~\ref{fig:asymdata}).

        In \cite{Viola2002Fast}, Viola and Jones pointed out the
        limitation of AdaBoost in the context of highly skewed example
        distribution and proposed a new variant of AdaBoost called
        AsymBoost which is experimentally shown to give a significant
        performance improvement over conventional boosting. In brief,
        the sample weights were updated before each round of boosting
        with the extra exponential term which causes the algorithm to
        gradually pay more attention to positive samples in each round
        of boosting.
        Our scheme based on LDA's class-separability can be considered
        as an alternative classifier to AsymBoost that also takes
        asymmetry information into consideration.

\subsection{Boosted Greedy Sparse Linear Discriminant Analysis}  
\label{sec:bgslda}

        Before we introduce the concept of BGSLDA, we present a brief
        explanation of boosting algorithms.  Boosting is one of the
        most popular learning algorithms. It was originally designed
        for classification problems. It combines the output of many
        weak classifiers to produce a single strong learner. Weak
        classifier is defined as a classifier with accuracy on the
        training set greater than average. There exist many variants
        of boosting algorithms, \eg, AdaBoost (minimizing the 
        exponential loss), GentleBoost (fitting regression function by weighted
        least square methods), LogitBoost (minimizing the logistic regression 
        cost function) \cite{Friedman2000Additive}, LPBoost
        (minimizing the Hinge loss)
        \cite{Demiriz2002LPBoost,Leskovec2003LPBoost}, \etc. 
        All of them have an identical
        property of sample re-weighting and weighted majority vote.
        One of the wildly used boosting algorithm is AdaBoost
        \cite{Schapire1999Boosting}. AdaBoost is a greedy algorithm
        that constructs an additive combination of weak classifiers
        such that the exponential loss 
        \[ 
              L(y,F( \bx )) = \exp ( -y F( \bx ) ) 
        \]
              is minimized. Here $\bx$ is the labeled training examples
        and $y$ is its label; $F( \bx)$ is the final decision function
        which outputs the decided class label. Each training sample
        receives a weight $u_i$ that determines its significance for
        training the next weak classifier. In each boosting iteration,
        the value of $\alpha_t$ is computed and the sample weights are
        updated according to the exponential rule. AdaBoost then
        selects a new hypothesis $h(\cdot)$ that best classifies
        updated training samples with minimal classification error
        $e$. The final decision rule $F(\cdot)$ is a linear
        combination of the selected weak classifiers weighted by their
        coefficients $\alpha_t$. The classifier decision is given by
        the sign of the linear combination
 \[ F( \bx ) =
                 \sign \Bigl(
                       \sum_{t=1}^{N_w} \alpha_{t}h_{t}(\bx)
                       \Bigr), \]         
where $\alpha_t$ is a weight coefficient; $h_t (\cdot)$ is a weak
learner and $ N_w$ is the number of weak classifiers. The expression
of the above equation is similar to an expression used in
dimensionality reduction where $F(\bx)$ can be considered as the
result of linearly projecting the random vector $\bF_i$ onto a one
dimensional space along the direction of $\boldsymbol \alpha$.

        In previous section, we have introduced the concept of GSLDA
        in the domain of object detection. However, decision stumps
        used in GSLDA algorithm are learned {\em only once} to save
        computation time. In other words, once learned, an optimal
        threshold, which gives smallest classification error on the
        training set, remains unchanged during GSLDA training.  This
        speeds up the training process as also shown in forward
        feature selection of \cite{Wu2008Fast}. However, it limits the
        number of decision stumps available for GSLDA classifier to
        choose from. As a result, GSLDA algorithm fails to perform at
        its best. In order to achieve the best performance from the
        GSLDA classifier, we propose to extend decision stumps used in
        GSLDA training with sample re-weighting techniques used in
        boosting methods. In other words, each training sample
        receives a weight and the new set of decision stumps are
        trained according to these sample weights. The objective
        criterion used to select the best decision stump is similar to
        the one applied in step ($1$) in Algorithm~\ref{ALG:GSLDA}.
        Note that step ($3$) in Algorithm~\ref{ALG:BGSLDA} is
        introduced in order to speed up the GSLDA training process. In
        brief, we remove decision stumps with weighted error larger
        than $e_k + \varepsilon$ where $e_k = \frac{1}{2} -
        \frac{1}{2}\beta_k$, $\beta_k = \max{(\sum_{i=1}^{N} u_i y_i
        h_t(x_i))}$ and $N$ is the number of samples, $y_i$ is the
        class label of sample $x_i$, $h_t(x_i)$ is the prediction of
        the training data $x_i$ using weak classifier $h_t$. 
        The condition used here has connection with
        the dual constraint of the soft
        margin LPBoost
        \cite{Demiriz2002LPBoost}. The dual objective of LPBoost
        minimizes $\beta$ subject to the constraints
        \[
        \textstyle \sum\nolimits_{i=1}^{N} u_i y_i h_t(x_i) \leq \beta, \forany
        t,
        \]
        and 
        \[
        \textstyle \sum\nolimits_{i=1}^{N} u_i = 1, 0 \leq u_i \leq {\rm
        const}, \forany i.
        \]
        As a result, the sample weights $u_i$ is
        the most pessimistic one. We choose decision stumps with
        weighted error smaller than $e_k + \varepsilon$. These
        decision stumps are the ones that perform best under the most
        pessimistic condition.

        Given the set of decision stumps, GSLDA selects the stump that
        results in maximum class separation (step ($4$)). The sample
        weights can be updated using different boosting algorithm
        (step ($5$)). In our experiments, we use AdaBoost
        \cite{Viola2004Robust} re-weighting scheme (BGSLDA - scheme
        $1$).
\begin{equation}
  \label{EQ:adaboost_reweight}
  D_{i}^{(t+1)} = \frac{ u_{i}^{(t)} \exp(-\alpha_t y_i h_t(\bx_i)) }
  {   Z^{(t+1)}  },
\end{equation}
with 
\[
  Z^{(t+1)} =   
{\sum_i u_{i}^{(t)} \exp(-\alpha_t y_i h_t(\bx_i)) }.
\]
Here $\alpha_t = \log((1-e_t)/(e_t))$ and $e_t$ is the weighted error.
We also use AsymBoost \cite{Viola2002Fast} re-weighting scheme (BGSLDA - scheme $2$). 
\begin{equation}
  \label{EQ:asymboost_reweight}
  u_{i}^{(t+1)} = \frac{ u_{i}^{(t)} \exp(-\alpha_t y_i
  h_t(\bx_i)) \exp(y_i \log \sqrt{k}) } {Z^{(t+1)}},
  \end{equation}
  with 
  \[
  Z^{(t+1)} = 
  {\sum_i u_{i}^{(t)} \exp(-\alpha_t
  y_i h_t(\bx_i)) \exp(y_i \log \sqrt{k}) }.
\]
Since BGSLDA based object detection framework has the same
input/output as GSLDA based detection framework, we replace lines $2 -
10$ in Algorithm~\ref{ALG:GSLDA} with Algorithm~\ref{ALG:BGSLDA}.

%
%
%
%
%
%
%

\begin{algorithm}[t]
\caption{The training algorithm for building a cascade of BGSLDA object detector.}
\begin{algorithmic}
\small{

\While{ $  \ftarget < F_i $ }
{
$ i = i + 1$;\\
$ f_i = 1 $;\\
   \While{$ f_i > \fmax$ }
      { 
          1. 
              Normalize sample weights $ \bu $;
          \\2. 
          Train weak classifiers $h(\cdot)$
          (\eg, decision stumps by finding an optimal
            threshold $ \theta $)  
            using the training set and sample weights;
          \\3.
          Remove those weak classifiers with weighted error larger than 
             $ e_k + \varepsilon$
            (section~\ref{sec:bgslda});
         \\4. 
         Add the weak classifier whose output yields the maximum class
             separation;
         \\5. 
         Update sample weights $ \bu $ in the AdaBoost manner
         (Eq.~\eqref{EQ:adaboost_reweight})
            or AsymBoost manner (Eq.~\eqref{EQ:asymboost_reweight});
         \\6.
         Lower threshold such that $\dmin$ holds;
         \\7.
         Update $f_i$ using this threshold;
      }
      $ D_{i+1} = D_i \times \dmin $;
        
      $ F_{i+1} = F_i \times f_i $; and
      remove those correctly classified negative samples from the training
      set;\\
      \If{ $\ftarget < F_i $ }
      {Evaluate the current cascaded classifier on the negative images
      and add misclassified samples into the negative training set;}
}

} 
\end{algorithmic}
\label{ALG:BGSLDA}
\end{algorithm}

\subsection{Training Time Complexity of BGSLDA}
\label{sec:complexity}

In order to analyze the complexity of the proposed system, we need to
analyze the complexity of boosting and GSLDA training. Let the number
of training samples in each cascade layer be $N$. For boosting,
finding the optimal threshold of each feature needs $O(N\log{N})$.
Assume that the size of the feature set is $M$ and the number of weak
classifiers to be selected is $T$. The time complexity for training
boosting classifier is $O(MTN\log{N})$. The time complexity for GSLDA
forward pass is $O(NMT + MT^3)$. $O(N)$ is the time complexity for
finding mean and variance of each features. $O(T^2)$ is the time
complexity for calculating correlation for each feature. Since, we
have $M$ features and the number of weak classifiers to be selected is
$T$, the total time for complexity for GSLDA is $O(NMT + MT^3)$.
Hence, the total time complexity is 
$O(\underbrace{MTN\log{N}}_{\text{weak classifier}} +
\underbrace{NMT + MT^3}_{\text{GSLDA}})$. Since, $T$ is often small
(less than $200$) 
in cascaded structure, the term $O(MTN\log{N})$ often
dominates. In other words, most of the computation time is spent on
training weak classifiers. 

\section{Experiments}
\label{sec:exp}

        This  section is organized as follows. The datasets used in
        this experiment, including how the performance is analyzed,
        are described. Experiments and the parameters used are then
        discussed.  Finally, experimental results and analysis of
        different techniques are presented.

	\subsection{Face Detection with the GSLDA Classifier}
	\label{sec:GSLDA}

        Due to its efficiency, Haar-like rectangle features
        \cite{Viola2004Robust} have become a popular choice as image
        features in the context of face detection. Similar to the work
        in \cite{Viola2004Robust}, the weak learning algorithm known
        as decision stump and Haar-like rectangle features are used
        here due to their simplicity and efficiency. 
%
%
%
%
        The following experiments compare AdaBoost and GSLDA learning
        algorithms in their performances in the domain of face
        detection. For fast AdaBoost training of Haar-like rectangle
        features, we apply the pre-computing technique similar to
        \cite{Wu2008Fast}.

  \subsubsection{Performances on Single-node Classifiers}
  \label{sec:onestage_gslda}
  
\begin{figure}[t]
            \centering
\includegraphics[width=0.03\textwidth]{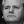}
\includegraphics[width=0.03\textwidth]{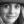}
\includegraphics[width=0.03\textwidth]{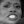}
\includegraphics[width=0.03\textwidth]{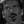}
\includegraphics[width=0.03\textwidth]{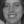}
\includegraphics[width=0.03\textwidth]{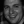}
\includegraphics[width=0.03\textwidth]{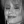}
\includegraphics[width=0.03\textwidth]{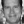}
\includegraphics[width=0.03\textwidth]{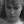}
\includegraphics[width=0.03\textwidth]{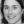}
\includegraphics[width=0.03\textwidth]{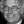}
\includegraphics[width=0.03\textwidth]{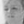}
\includegraphics[width=0.03\textwidth]{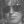}
\includegraphics[width=0.03\textwidth]{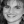}
\includegraphics[width=0.03\textwidth]{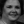}
\includegraphics[width=0.03\textwidth]{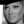}
\includegraphics[width=0.03\textwidth]{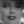}
\includegraphics[width=0.03\textwidth]{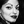}
\includegraphics[width=0.03\textwidth]{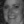}
\includegraphics[width=0.03\textwidth]{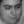}
\includegraphics[width=0.03\textwidth]{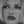}
\includegraphics[width=0.03\textwidth]{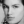}
\includegraphics[width=0.03\textwidth]{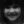}
\includegraphics[width=0.03\textwidth]{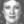}
\includegraphics[width=0.03\textwidth]{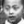}
\includegraphics[width=0.03\textwidth]{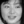}
\includegraphics[width=0.03\textwidth]{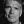}
\includegraphics[width=0.03\textwidth]{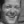}
\includegraphics[width=0.03\textwidth]{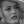}
\includegraphics[width=0.03\textwidth]{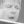}
\includegraphics[width=0.03\textwidth]{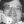}
\includegraphics[width=0.03\textwidth]{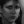}
\includegraphics[width=0.03\textwidth]{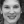}
\includegraphics[width=0.03\textwidth]{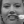}
\includegraphics[width=0.03\textwidth]{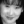}
\includegraphics[width=0.03\textwidth]{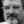}
\includegraphics[width=0.03\textwidth]{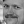}
\includegraphics[width=0.03\textwidth]{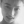}
\includegraphics[width=0.03\textwidth]{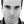}
\includegraphics[width=0.03\textwidth]{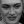}
\includegraphics[width=0.03\textwidth]{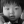}
\includegraphics[width=0.03\textwidth]{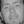}
\includegraphics[width=0.03\textwidth]{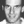}
\includegraphics[width=0.03\textwidth]{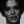}
\includegraphics[width=0.03\textwidth]{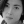}
\includegraphics[width=0.03\textwidth]{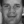}
\includegraphics[width=0.03\textwidth]{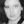}
\includegraphics[width=0.03\textwidth]{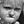}
\includegraphics[width=0.03\textwidth]{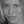}
\includegraphics[width=0.03\textwidth]{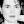}
\includegraphics[width=0.03\textwidth]{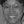}
\includegraphics[width=0.03\textwidth]{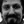}
\includegraphics[width=0.03\textwidth]{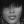}
\includegraphics[width=0.03\textwidth]{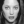}
\includegraphics[width=0.03\textwidth]{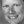}
\includegraphics[width=0.03\textwidth]{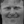}
\includegraphics[width=0.03\textwidth]{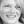}
\includegraphics[width=0.03\textwidth]{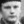}
\includegraphics[width=0.03\textwidth]{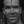}
\includegraphics[width=0.03\textwidth]{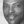}
\includegraphics[width=0.03\textwidth]{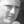}
\includegraphics[width=0.03\textwidth]{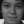}
\includegraphics[width=0.03\textwidth]{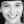}
\includegraphics[width=0.03\textwidth]{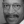}
\includegraphics[width=0.03\textwidth]{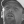}
            \caption{A random sample of face  images for training.}
            \label{FIG:training_faces}
        \end{figure}

  This experiment compares single strong classifier learned using
  AdaBoost and GSLDA algorithms in their classification performance.
  The datasets consist of three training sets and two test sets. Each
  training set contains $2,000$ face examples and $2,000$ non-face
  examples (Table~\ref{tab:1}). 
        The dataset consists of $10,000$ mirrored faces. The faces
        were cropped and rescaled to images of size $24 \times 24$
        pixels. For non-face examples, we randomly selected $10,000$
        random non-face patches from non-face images obtained from the
        internet. 
        Fig.~\ref{FIG:training_faces}
        shows a random sample of face training images. 

\begin{table}[]
    \caption{The size of training and test sets used on the single node classifier.}
	\begin{center}
	\begin{tabular}{l|ccc}
	\hline
	$  $\#$ $ & data splits  &  faces$/$split & non-faces$/$split\\
		\hline
        \hline
			Train   &  $3$ & $2000$ & $2000$  \\
			Test    &  $2$ & $2000$ & $2000$  \\
		\hline
	\end{tabular}
	\end{center}
\label{tab:1}
\end{table}

\begin{figure*}[t!]
			\begin{center}
				\subfigure[]{\includegraphics[width=0.42\textwidth, clip]{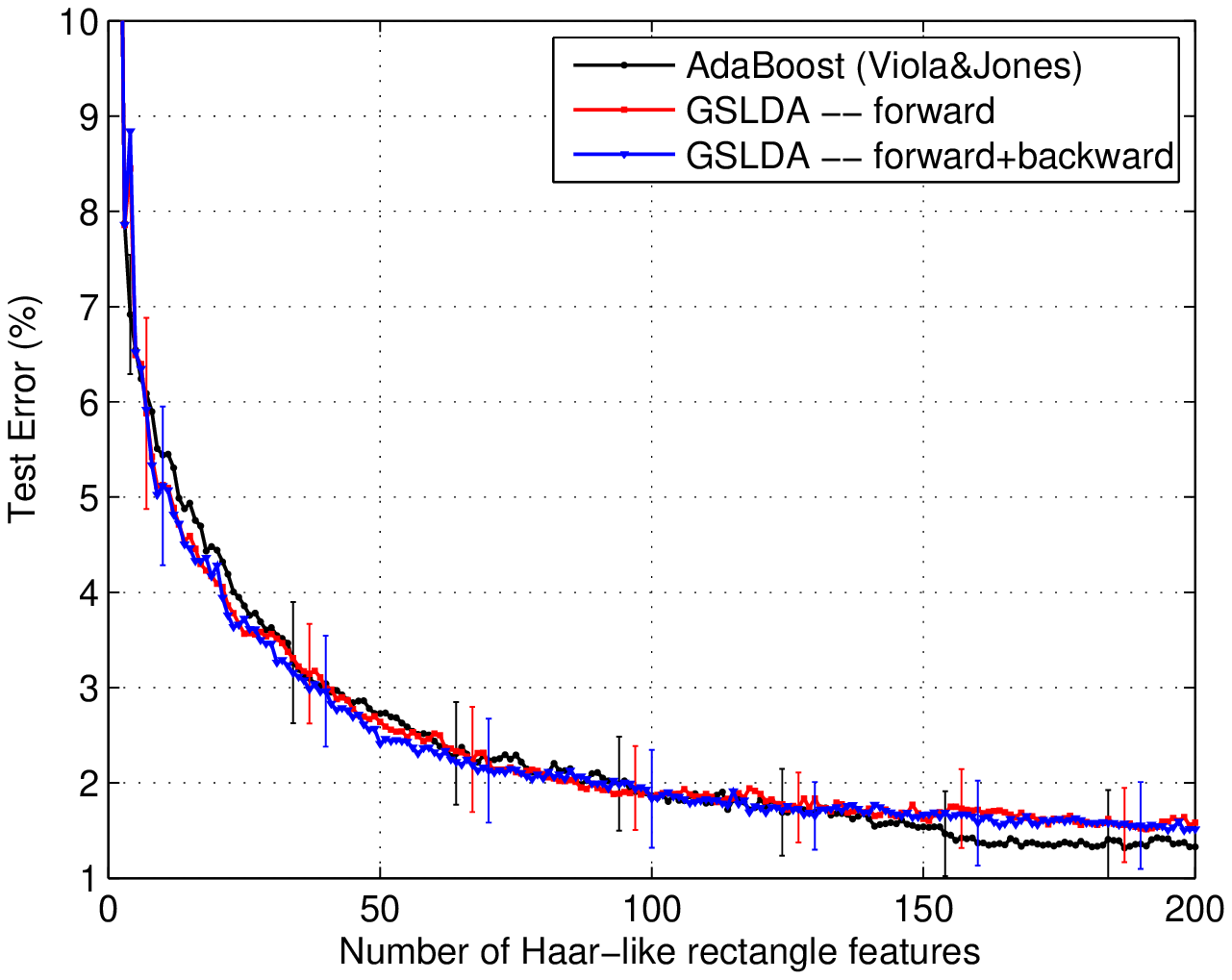}
					\label{fig:one_a}
				}
				\subfigure[]{\includegraphics[width=0.42\textwidth,clip]{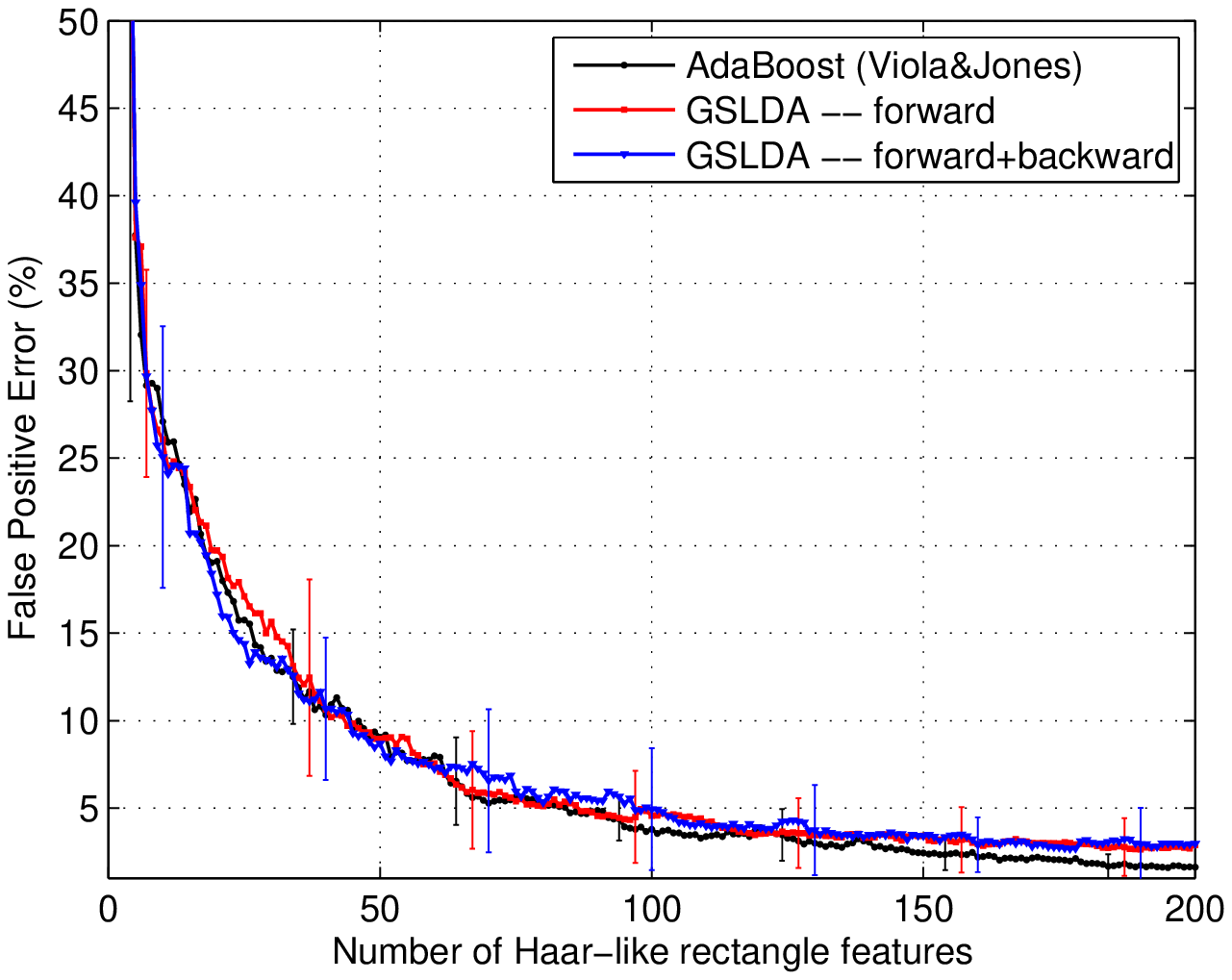}
					\label{fig:one_b}					
			  }
				\subfigure[]{\includegraphics[width=0.42\textwidth, clip]{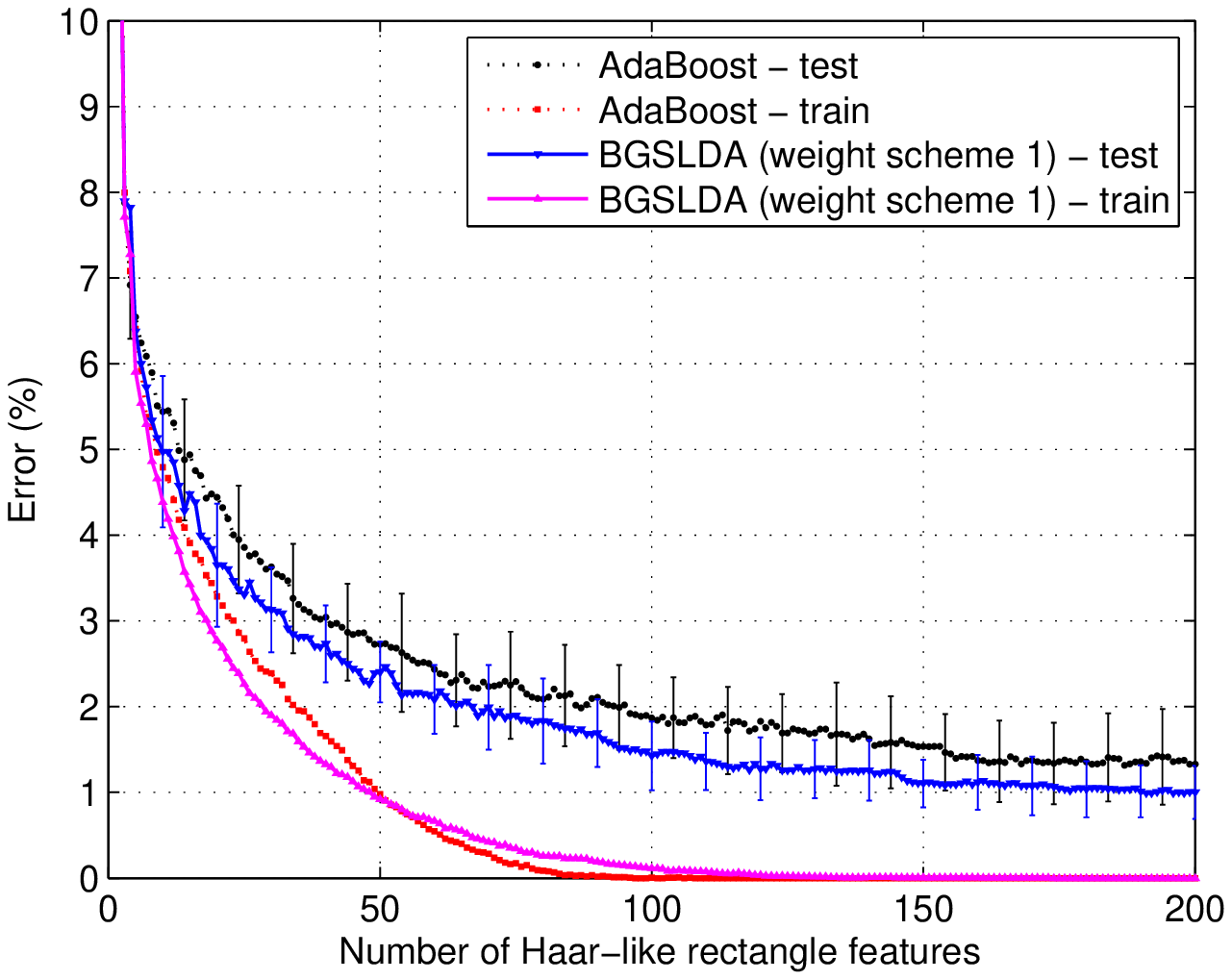}
					\label{fig:one_c}
			  }
				\subfigure[]{\includegraphics[width=0.42\textwidth, clip]{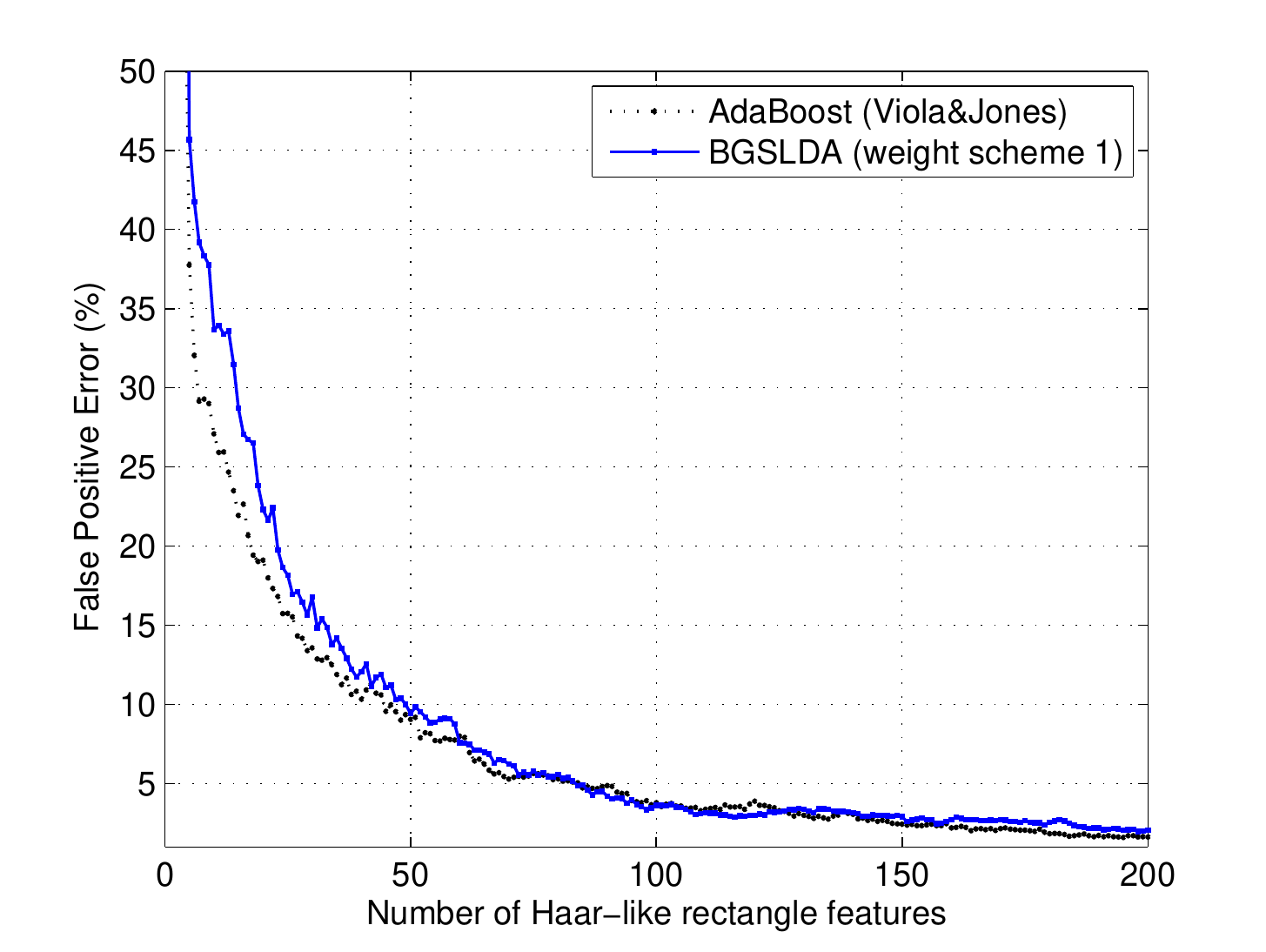}
					\label{fig:one_d}
			  }
			\end{center}
		\caption{See text for details (best viewed in color). 
			(a) Comparison of test error rates between GSLDA and AdaBoost.
 		  (b) Comparison of false alarm rates on test set between GSLDA
 		      and AdaBoost. The detection rate on the validated face set is fixed at $99\%$.
 		  (c) Comparison of train and test error rates between BGSLDA (scheme $1$) and AdaBoost.
		  (d) Comparison of false alarm rates on test set between BGSLDA (scheme $1$) and AdaBoost.
		  }
		\label{fig:onestage}
	\end{figure*}

        For each experiment, three different classifiers are
        generated, each by selecting two out of the three training
        sets and the remaining training set for validation. 
        The performance is measured by two different curves:- the test
        error rate and the classifier learning goal (the false alarm
        error rate on test set given that the detection rate on the
        validation set is fixed at $99\%$). A $95\%$ confidence
        interval of the true mean error rate is given by the
        t-distribution. In this experiment, we test two different
        approaches of GSLDA: forward-pass GSLDA and dual-pass
        (forward+backward) GSLDA. 
        The results are shown in
        Fig.~\ref{fig:onestage}. The following observations can be
        made from these curves. Having the same number of learned
        Haar-like rectangle features, GSLDA achieves a comparable
        error rate to AdaBoost on test sets (Fig.~\ref{fig:one_a}).
        GSLDA seems to perform slightly better with less number of
        Haar-like features ($<100$) while AdaBoost seems to perform
        slightly better with more Haar-like features ($>100$).
        However, both classifiers perform almost similarly within
        $95\%$ confidence interval of the true error rate. This
        indicates that features selected using GSLDA classifier are as
        meaningful as features selected using AdaBoost classifier.
        From the curve, GSLDA with bi-directional search yields better
        results than GSLDA with forward search only. 
        %
        %
        %
%
%
%
        Fig.~\ref{fig:one_b} shows the false positive error rate on
        test set.
%
%
%
From the figure, both GSLDA and AdaBoost achieve a comparable false positive error rate on test set. 

	\subsubsection{Performances on Cascades of Strong Classifiers}
	\label{sec:cascade_gslda}

%
%
        In this experiment, we used $5,000$ mirrored faces from
        previous experiment. The non-face samples used in each cascade
        layer are collected from false positives of the previous
        stages of the cascade (bootstrapping). The cascade training
        algorithm terminates when there are not enough negative
        samples to bootstrap. For fair evaluation, we trained both
        techniques with the same number of weak classifiers in each
        cascade. Note that since dual pass GSLDA (forward+backward
        search) yields better solutions than the forward search in the
        previous experiment, we use dual pass GSLDA classifier to
        train a cascade of face detectors.
%
%
        We tested our face detectors on the low resolution faces
        dataset, MIT+CMU frontal face test set. The complete set
        contains $130$ images with $507$ frontal faces. In this
        experiment, we set the scaling factor to $1.2$ and window
        shifting step to $1$. The technique used for merging
        overlapping windows is similar to \cite{Viola2004Robust}.
        Detections are considered true or false positives based on the
        area of overlap with ground truth bounding boxes. To be
        considered a correct detection, the area of overlap between
        the predicted bounding box and ground truth bounding box must
        exceed $50\%$. Multiple detections of the same face in an
        image are considered false detections.

\begin{figure*}[thb!]
			\begin{center}
				\subfigure[]{\includegraphics[width=0.42\textwidth, clip]{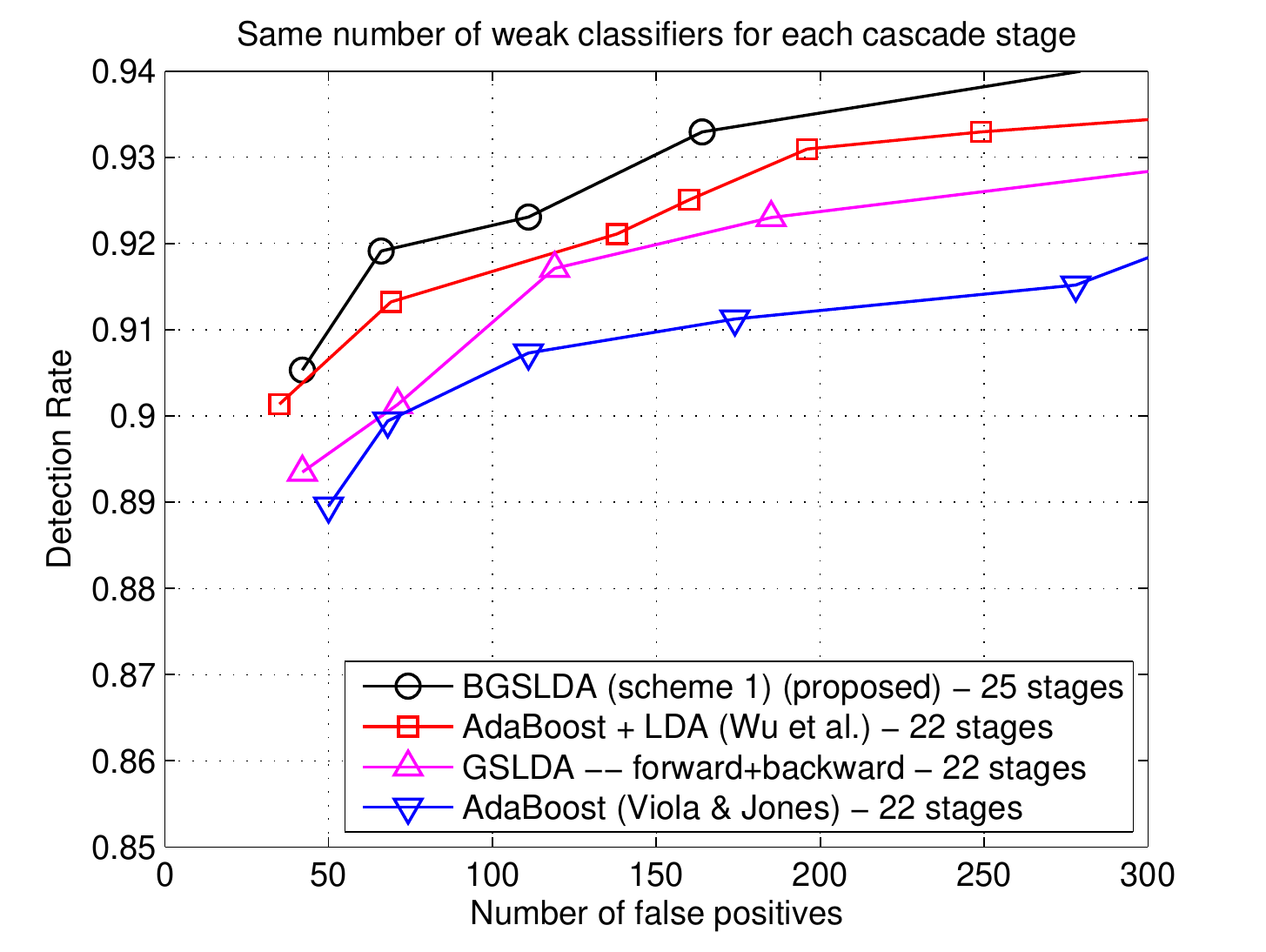}
					\label{fig:cascade_a}
				}
				\subfigure[]{\includegraphics[width=0.42\textwidth,clip]{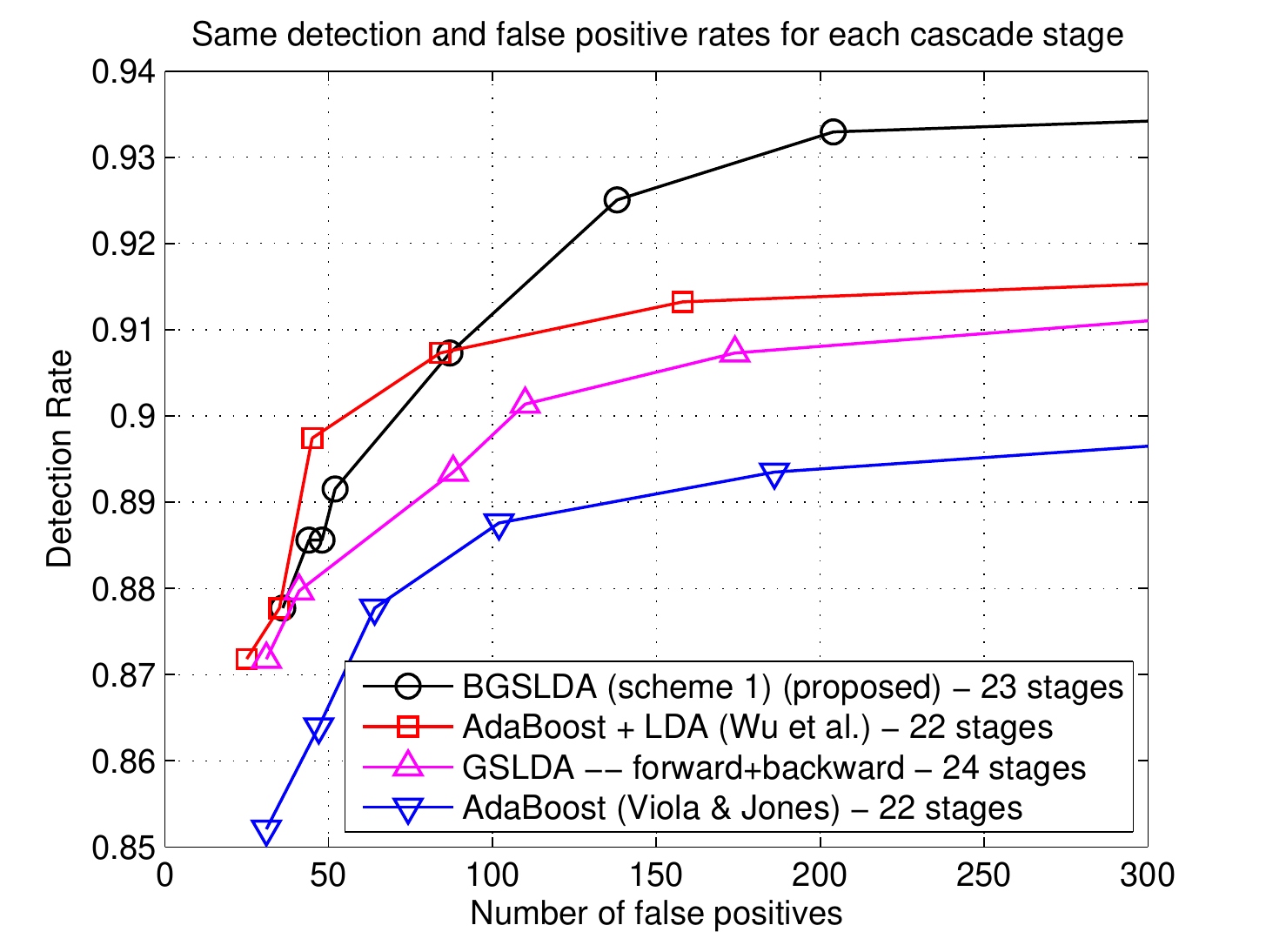}
					\label{fig:cascade_b}					
			  }
			\end{center}
		\caption{Comparison of ROC curves on the MIT+CMU face test set
			(a) with the same number of weak classifiers in each cascade stage
			    on AdaBoost and its variants.
 		  (b) with $99.5\%$ detection rate and $50\%$ false positive rate
 		      in each cascade stage on AdaBoost and its variants. 
 		      BGSLDA (scheme 1) corresponds to GSLDA classifier with decision
 		      stumps being re-weighted using AdaBoost scheme.
 		      	}
		\label{fig:cascade1}
	\end{figure*}

        Figs.~\ref{fig:cascade_a} and \ref{fig:cascade_b} show a
        comparison between the 
        Receiver Operating Characteristic
        (ROC) curves produced by GSLDA classifier
        and AdaBoost classifier. In Fig.~\ref{fig:cascade_a}, the
        number of weak classifiers in each cascade stage is
        predetermined while in Fig.~\ref{fig:cascade_b}, weak
        classifiers are added to the cascade until the predefined
        objective is met. 
%
%
        The ROC curves show that GSLDA classifier outperforms AdaBoost
        classifier at all false positive rates. We think that by
        adjusting the threshold to the AdaBoost classifier (in order
        to achieve high detection rates with moderate false positive
        rates), the performance of AdaBoost is no longer optimal. Our
        findings in this work are consistent with the experimental
        results reported in \cite{Viola2002Fast} and
        \cite{Wu2008Fast}. 
%
%
        \cite{Wu2008Fast} used LDA weights instead of weak
        classifiers' weights provided by AdaBoost algorithm.

        GSLDA not only performs better than AdaBoost but it is also
        much simpler. Weak classifiers learning (decision stumps) is
        performed only once for the given set of samples (unlike
        AdaBoost where weak classifiers have to be re-trained in each
        boosting iteration). GSLDA algorithm sequentially selects
        decision stump whose output yields the maximum eigenvalue. The
        process continues until the stopping criteria are met. Note
        that given the decision stumps selected by GSLDA, any linear
        classifiers can be used to calculate the weight coefficients.
        Based on our experiments, using linear SVM (maximizing the
        minimum margin) instead of LDA also gives a very similar
        result to our GSLDA detector. We believe that using one
        objective criterion for feature selection and another
        criterion for classifier construction would provide a
        classifier with more flexibility than using the same criterion
        to select feature and train weight coefficients. These
        findings open up many more possibilities in combining various
        feature selection techniques with many existing classification
        techniques. We believe that a better and faster object
        detector can be built with careful design and experiment.

\begin{figure}[t!]
	  \begin{center}

		\includegraphics[width=0.48\textwidth,clip]{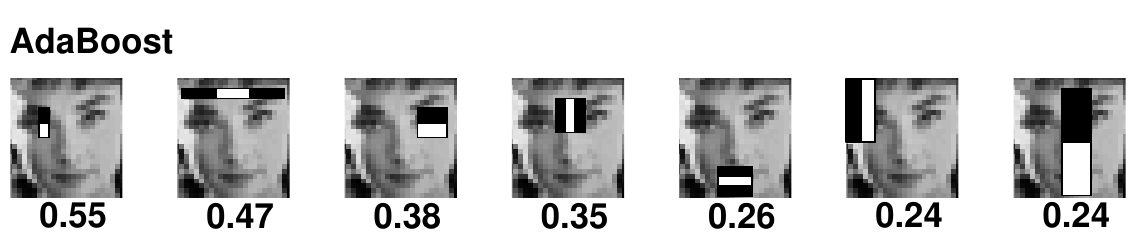}
		\includegraphics[width=0.48\textwidth,clip]{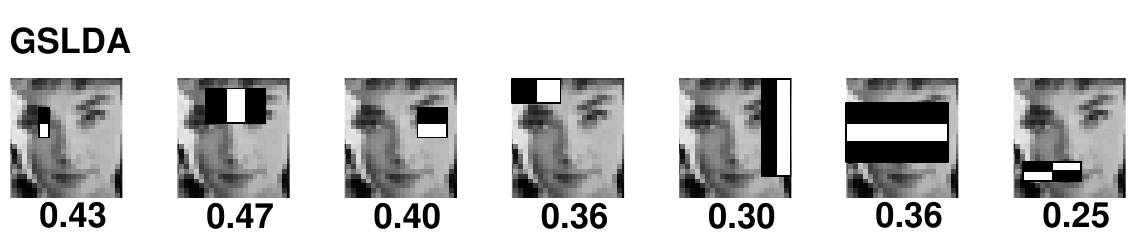}
	    \includegraphics[width=0.48\textwidth,clip]{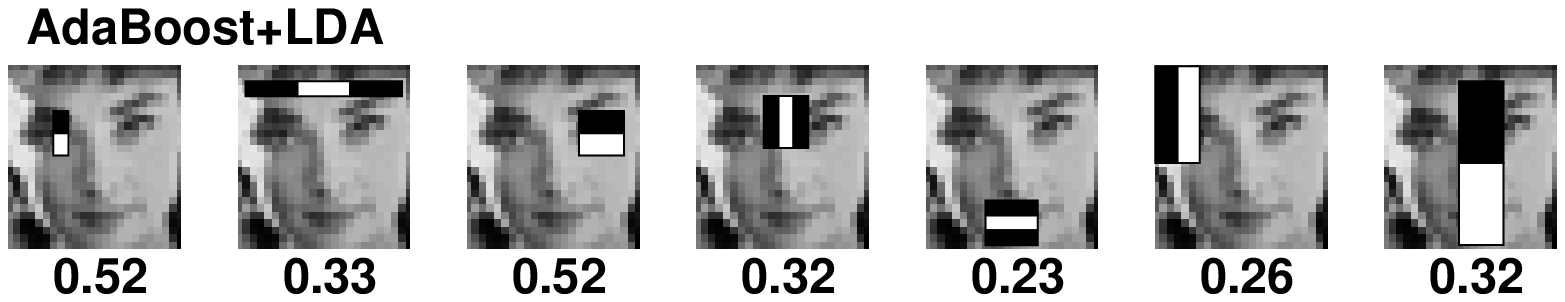}
		\includegraphics[width=0.48\textwidth,clip]{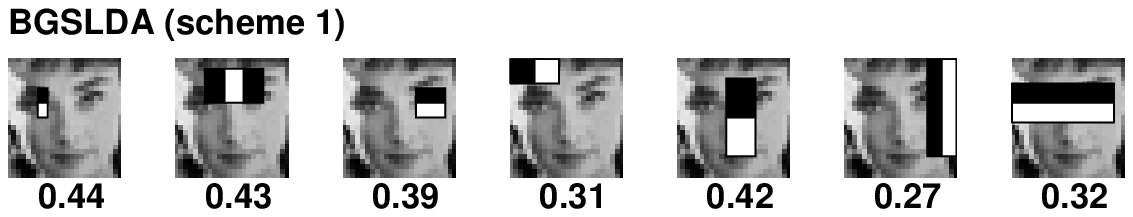}
		\end{center}
		\caption{
			      The first seven Haar-like rectangle features selected from
			      the first layer of the cascade. The value below each Haar-like
			      rectangle features indicates the normalized feature weight.
			      For AdaBoost, the value corresponds to the normalized
			      $\alpha$ where $\alpha$ is computed from 
			      $\log((1-e_t)/e_t)$ and
			      $e_t$ is the weighted error. For LDA,
			      the value corresponds to the normalized $\bw$ such that
			      for input vector $\bx$ and a class label $y$, $\bw^{\T}\bx$
			      leads to maximum separation between two classes.
			      }
	 \label{fig:faces}
  \end{figure}

        Haar-like rectangle features selected in the first cascade
        layer of both classifiers are shown in Fig.~\ref{fig:faces}.
        Note that both classifiers select Haar-like features which
        cover the area around the eyes and forehead.
        Table~\ref{tab:compare} compares the two cascaded classifiers
        in terms of the number of weak classifiers and the average
        number of Haar-like rectangle features evaluated per detection
        window.  Comparing GSLDA with AdaBoost, we found that GSLDA
        performance gain comes at the cost of a higher computation
        time. 
        This is not surprising since the number of decision stumps
        available for training GSLDA classifier is much smaller than
        the number of decision stumps used in training AdaBoost
        classifier. Hence, AdaBoost classifier can choose a more
        powerful/meaningful decision stump. Nevertheless, GSLDA
        classifier outperforms AdaBoost classifier. This indicates
        that the classifier trained to maximize class separation might
        be more suitable in the domain where the distribution of
        positive and negative samples is highly skewed.  In the next
        section, we conduct an experiment on BGSLDA.



	\begin{figure*}[thb!]
			\begin{center}
			
				\subfigure[]{\includegraphics[width=0.42\textwidth, clip]{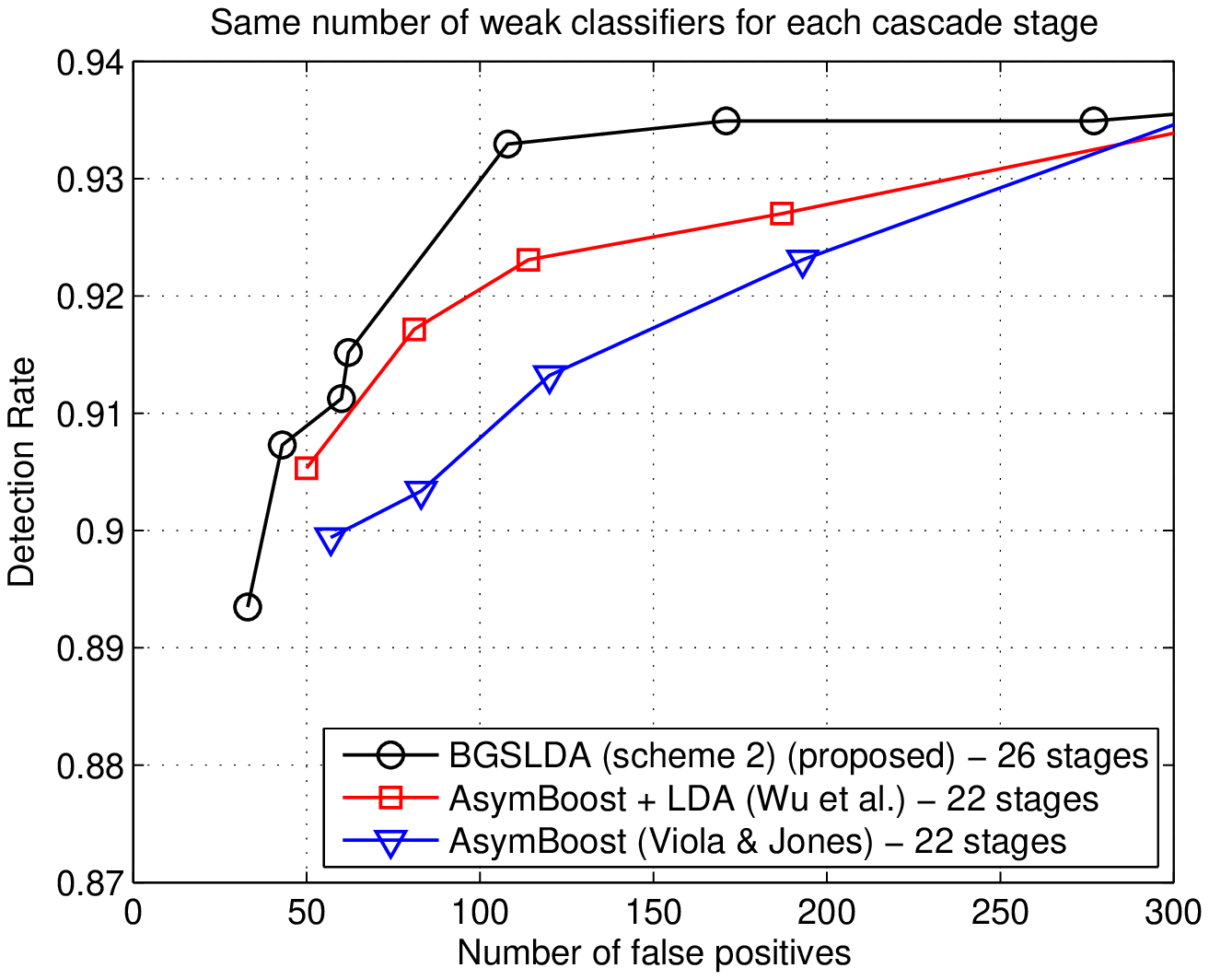}
					\label{fig:cascade_c}
			  }
				\subfigure[]{\includegraphics[width=0.42\textwidth, clip]{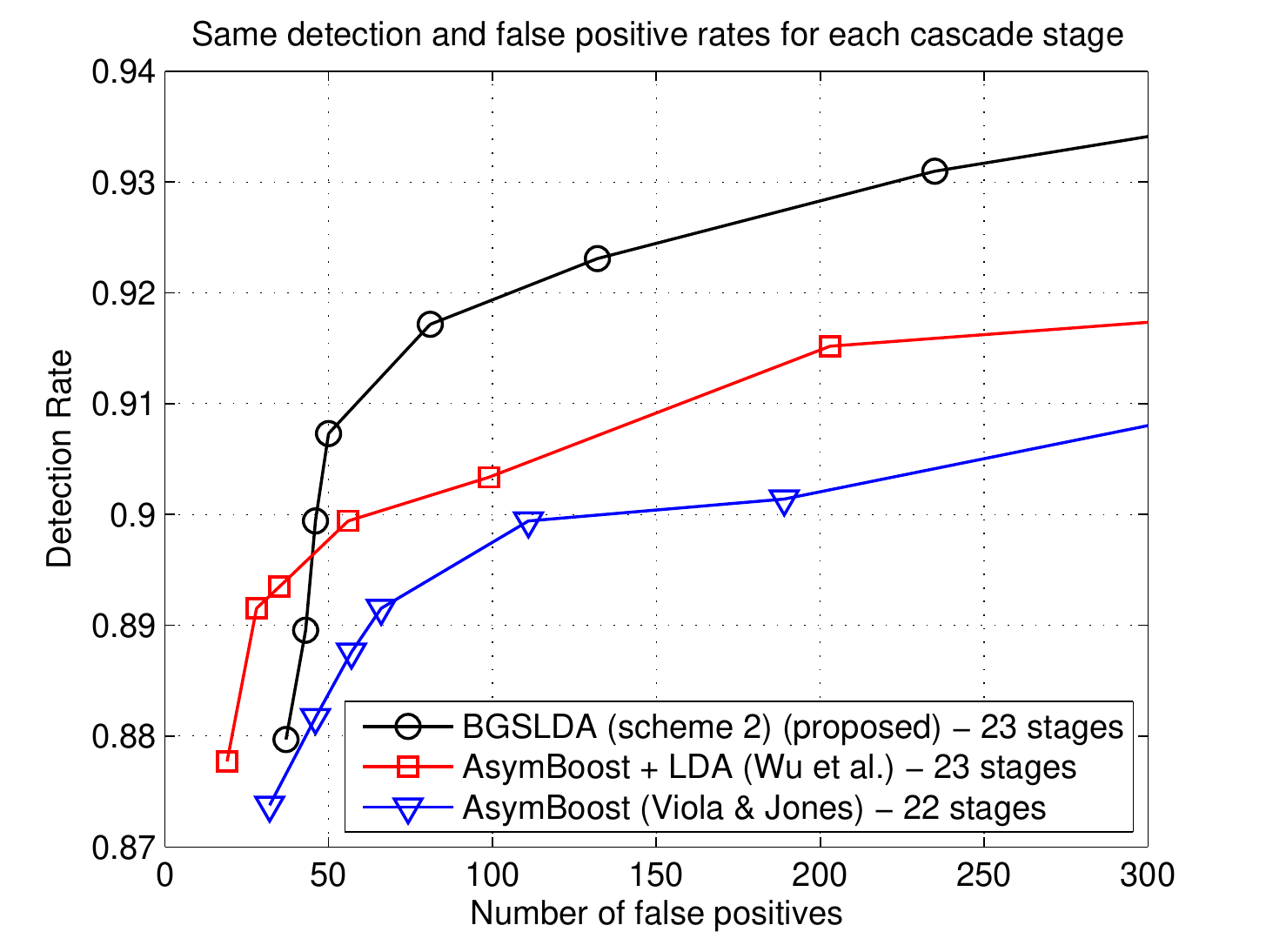}
					\label{fig:cascade_d}
			  }			
			\end{center}
		\caption{Comparison of ROC curves on the MIT+CMU face test set
			(a) with the same number of weak classifiers in each cascade stage
			    on AsymBoost and its variants.
 		  (b) with $99.5\%$ detection rate and $50\%$ false positive rate
 		      in each cascade stage on AsymBoost and its variants.
 		      BGSLDA (scheme 2) corresponds to GSLDA classifier with decision
 		      stumps being re-weighted using AsymBoost scheme.		}
		\label{fig:cascade2}
	\end{figure*}

        \subsection{Face Detection with BGSLDA classifiers}
	    \label{sec:BoostedGSLDA}

        The following experiments compare BGSLDA and different
        boosting learning algorithms in their performances for face
        detection. BGSLDA (weight scheme $1$) corresponds to GSLDA
        classifier with decision stumps being re-weighted using
        AdaBoost scheme while BGSLDA (weight scheme $2$) corresponds
        to GSLDA classifier with decision stumps being re-weighted
        using AsymBoost scheme (for highly skewed sample
        distributions). AsymBoost used in this experiment is from
        \cite{Viola2002Fast}. However, any asymmetric boosting
        approach can be applied here \eg
        \cite{Fan1999AdaCost,Leskovec2003LPBoost}.

  \subsubsection{Performances on Single Node Classifiers}
  \label{sec:onestage_boostedgslda}

The experimental setup is similar to the one described in previous section. The results are shown in figure~\ref{fig:onestage}. The following conclusions can be made from figure~\ref{fig:one_c}. Given the same number of weak classifiers, BGSLDA always achieves lower generalization error rate than AdaBoost. However, in terms of training error, AdaBoost achieves lower training error rate than BGSLDA. This is not surprising since AdaBoost has a faster convergence rate than BGSLDA. From the figure, AdaBoost only achieves lower training error rate than BGSLDA when the number of Haar-like rectangle features $> 50$. Fig.~\ref{fig:one_d} shows the false alarm error rate. 
The false positive error rate of both classifiers are quite similar.

  \subsubsection{Performances on Cascades of Strong Classifiers}
  \label{sec:cascade_boostedglda}

\begin{figure}[t]
        \centering
        	  \includegraphics[width=0.38\textwidth]{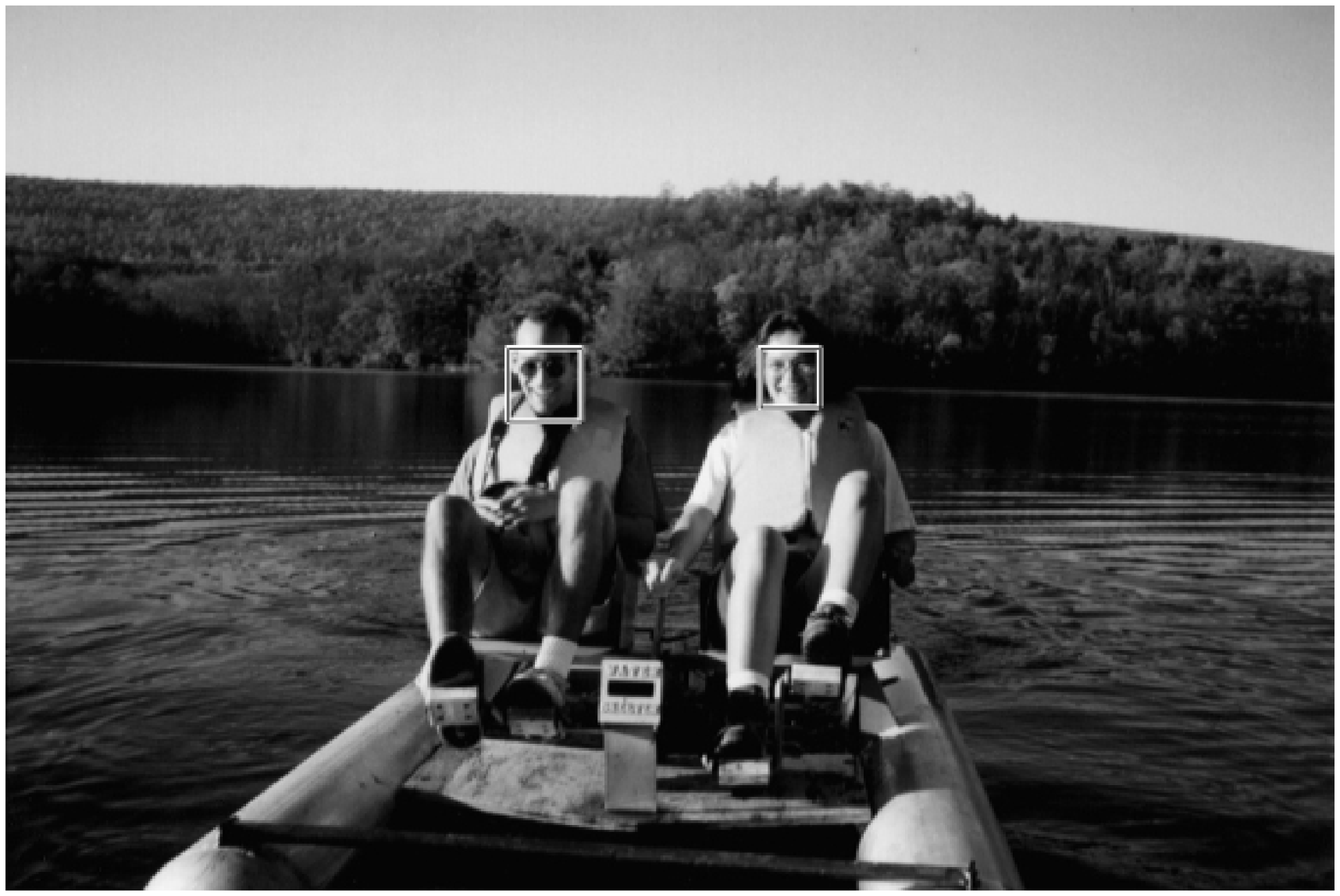}
              \includegraphics[width=0.38\textwidth]{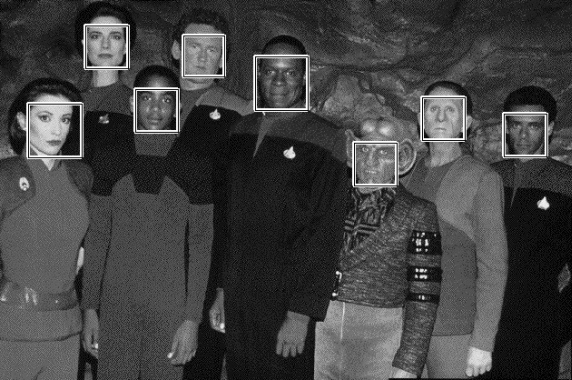}
              \includegraphics[width=0.38\textwidth]{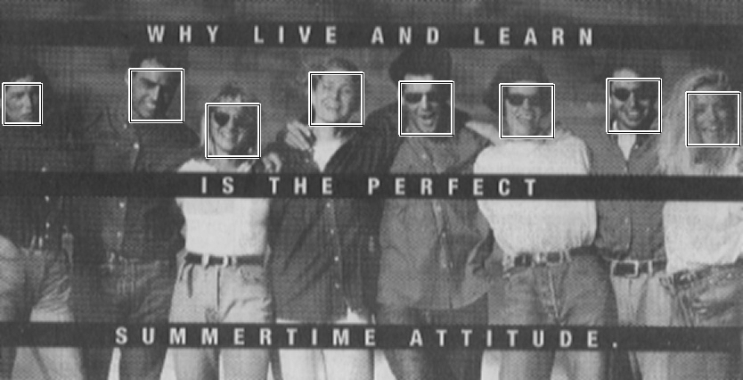}
              \includegraphics[width=0.38\textwidth]{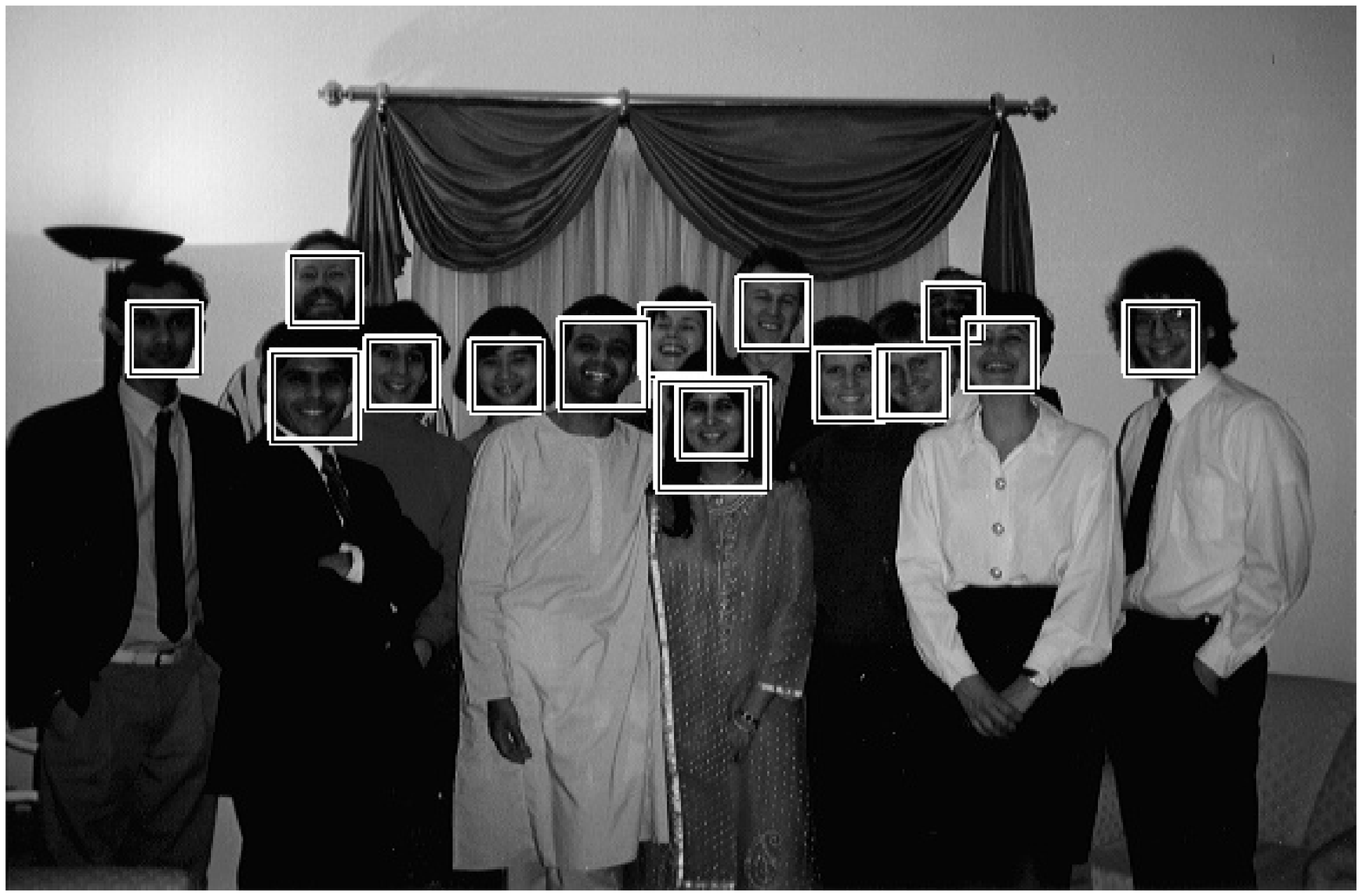}
        \caption{
		Face detection examples using the BGSLDA (scheme 1) detector 
        on the MIT+CMU test dataset.
		We set the scaling factor to $1.2$ and window shifting step to $1$ pixel. 
		The technique used for merging overlapping windows is similar to \cite{Viola2004Robust}.
        }  
        \label{FIG:face_results}
        \end{figure}

        The experimental setup and evaluation techniques used here are
        similar to the one described in
        Section~\ref{sec:onestage_gslda}. The results are shown in
        Fig.~\ref{fig:cascade1}. Fig.~\ref{fig:cascade_a} shows a
        comparison between the ROC curves produced by BGSLDA (scheme
        1) classifier and AdaBoost classifier trained with the same
        number of weak classifiers in each cascade. Both ROC curves
        show that the BGSLDA classifier outperforms both AdaBoost and
        AdaBoost+LDA \cite{Wu2008Fast}. Fig.~\ref{fig:cascade_b} shows
        a comparison between the ROC curves of different classifiers
        when the number of weak classifiers in each cascade stage is
        no longer predetermined. At each stage, weak classifiers are
        added until the predefined objective is met. Again, BGSLDA
        significantly outperforms other evaluated classifiers. 
        Fig.~\ref{FIG:face_results}
        demonstrates some face detection results on our BGSLDA (scheme 1) detector.

In the next experiment, we compare the performance of BGSLDA (scheme
2) with other classifiers using asymmetric weight updating rule
\cite{Viola2002Fast}. In other words, the asymmetric multiplier
$\exp(\frac{1}{N} y_i \log{\sqrt{k}})$ is applied to every sample
before each round of weak classifier training. 
        The results are shown in Fig.~\ref{fig:cascade2}.
        Fig.~\ref{fig:cascade_c} shows a comparison between the ROC
        curves trained with the same number of weak classifiers in
        each cascade stage. Fig.~\ref{fig:cascade_d} shows the ROC
        curves trained with $99.5\%$ detection rate and $50\%$ false
        positive rate criteria. From both figures, BGSLDA (scheme 2)
        classifier outperforms other classifiers evaluated. BGSLDA
        (scheme 2) classifier also outperforms BGSLDA (scheme 1)
        classifier. This indicates that asymmetric loss might be more
        suitable in domains where the distribution of positive
        examples and negative examples is highly imbalanced. Note that
        the performance gain between BGSLDA (scheme 1) and BGSLDA
        (scheme 2) is quite small compared with the performance gain
        between AdaBoost and AsymBoost. Since, LDA takes the number of
        samples of each class into consideration when solving the
        optimization problem, we believe this reduces the performance
        gap between BGSLDA (scheme 1) and BGSLDA (scheme 2).


        Table~\ref{tab:compare} indicates that our BGSLDA (scheme
        1) classifier performs at a speed comparable to AdaBoost
        classifier.  However, compared with AdaBoost+LDA, the
        performance gain of BGSLDA comes at the slightly higher cost
        in computation time.
        In terms of cascade training time, on a desktop with an Intel
        Core\texttrademark~$2$ Duo CPU T$7300$ with $4$GB RAM, the
        total training time is less than one day.

\begin{figure}
	  \begin{center}
		\includegraphics[width=0.40\textwidth,clip]{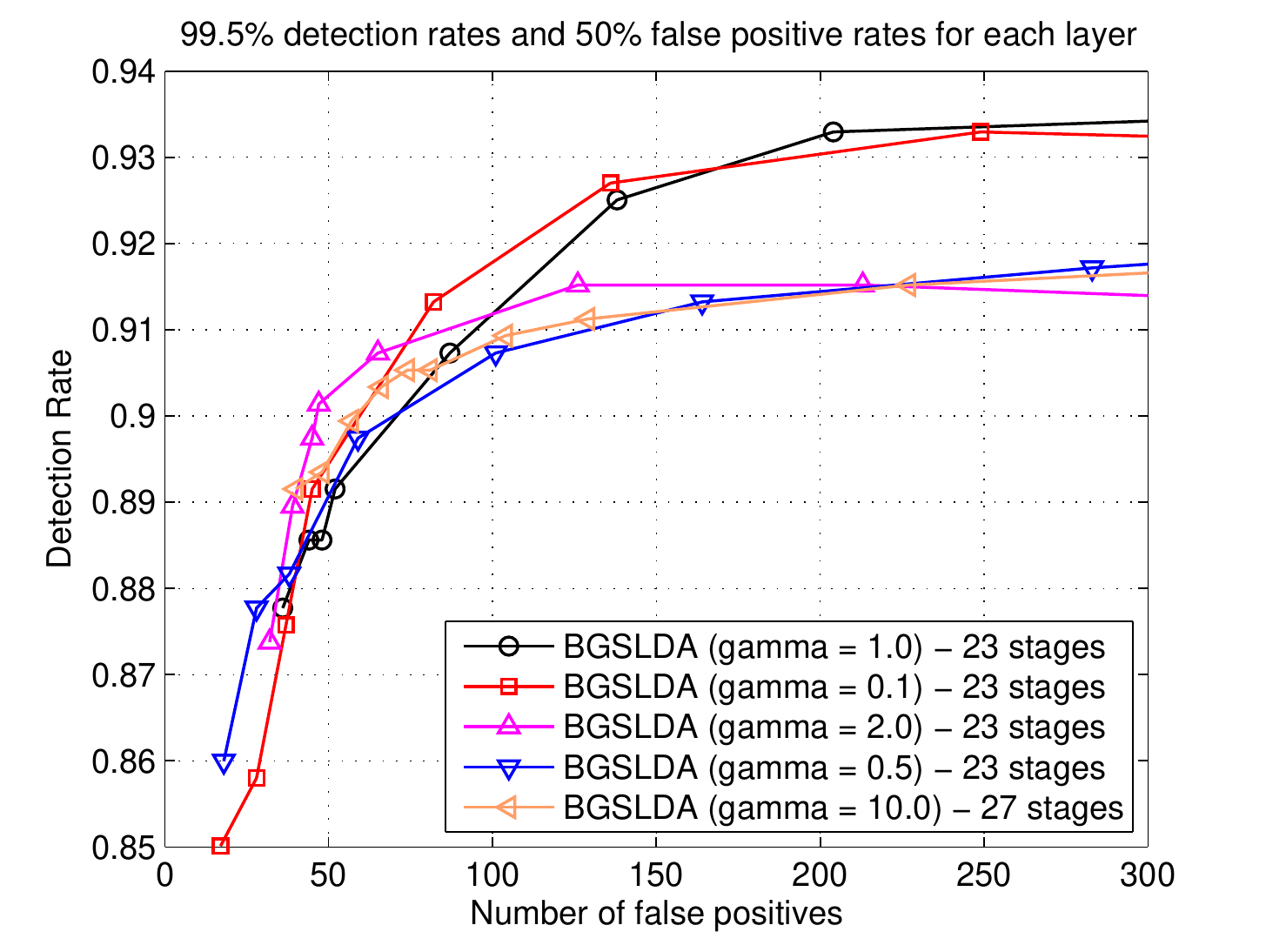}
		\end{center}
		\caption{
		         Comparison of ROC curves with different values of 
                 $ \gamma $ in \eqref{EQ:gamma}.
		}
	 \label{fig:lda1}
  \end{figure}

        As mentioned in \cite{Cooke2002Optimal}, a more general
        technique for generating discriminating hyperplanes is to
        define the total within-class covariance matrix as 
        \begin{align}
        S_w  &= \textstyle \sum_{x_i \in C_{1}}
        {(x_i - \mu_{1})(x_i - \mu_{1})^{\T}}  
                                          \notag \\ & 
        + \gamma \textstyle \sum_{x_i \in
        C_{2}}{(x_i - \mu_{2})(x_i - \mu_{2})^{\T}},
        \label{EQ:gamma}
\end{align}
        where $\mu_{1}$ is the mean of class $1$ and $\mu_{2}$ is the
        mean of class $2$. The weighting parameter $\gamma$ controls
        the weighted classification error. We have conducted an experiment
        on BGSLDA (scheme $1$) with different value of $\gamma$,
        namely $\gamma \in \left\{0.1,0.5,1.0,2.0,10.0\right\}$.
        All the other experiment settings remain the same as described in the
        previous section. 
        The results are
        shown in Fig.~\ref{fig:lda1}. 
        Based on ROC curves, it can be seen that 
        {\em all
        configurations of BGSLDA classifiers outperform 
        AdaBoost classifier
        at all false positive rates}.
        Setting $\gamma =1$ 
        gives the highest detection rates when the number of false
        positives is larger than $ 200$. 
        Setting $\gamma = 0.5$ performs best when the
        number of false positives is very small.

  

	\begin{table*}[bt]
	 \caption{Comparison of number of weak classifiers.  The number of
        cascade stages and total weak classifiers were obtained from
        the classifiers trained to achieve a detection rate of
        $99.5\%$ and the maximum false positive rate of $50\%$ in each
        cascade layer. The average number of Haar-like rectangles
        evaluated was obtained from evaluating the trained classifiers
        on MIT+CMU face test set.
        }
		\begin{center}
		\begin{tabular}{l|c|c|c}
		\hline
		method &    number of  stages &  total number  of weak
        classifiers   & average number of Haar features evaluated\\
		\hline\hline
		AdaBoost \cite{Viola2004Robust}      & $ 22$ & $ 1771$ & $ 23.9$     \\
		AdaBoost+LDA \cite{Wu2008Fast}     & $ 22$ & $ 1436$	& $ 22.3$	   \\
		GSLDA                                & $ 24$ & $ 2985$	& $ 36.0$	   \\
		BGSLDA (scheme 1)                     & $ 23$ & $ 1696$	 & $ 24.2$	 \\
		AsymBoost \cite{Viola2002Fast}       & $ 22$ & $ 1650$	& $ 22.6$	   \\
		AsymBoost+LDA \cite{Wu2008Fast}    & $ 22$ & $ 1542$	& $ 21.5$	   \\
		BGSLDA (scheme 2)                    & $ 23$ & $ 1621$	& $ 24.9$	   \\
	  \hline
	  \end{tabular}
		\end{center}
	  \label{tab:compare}
  \end{table*}

  \subsection{Pedestrian Detection with GSLDA and BGSLDA Classifiers}
        \label{sec:ped}
        
        In this section, we apply the proposed algorithm to pedestrian detection,
        which is considered a more difficult problem than face detection.

    \subsubsection{Pedestrian Detection on the Daimler-Chrsyler dataset
    with Haar-like Features}
    
        \begin{figure}[t!]
            \centering
                \includegraphics[width=0.42\textwidth]{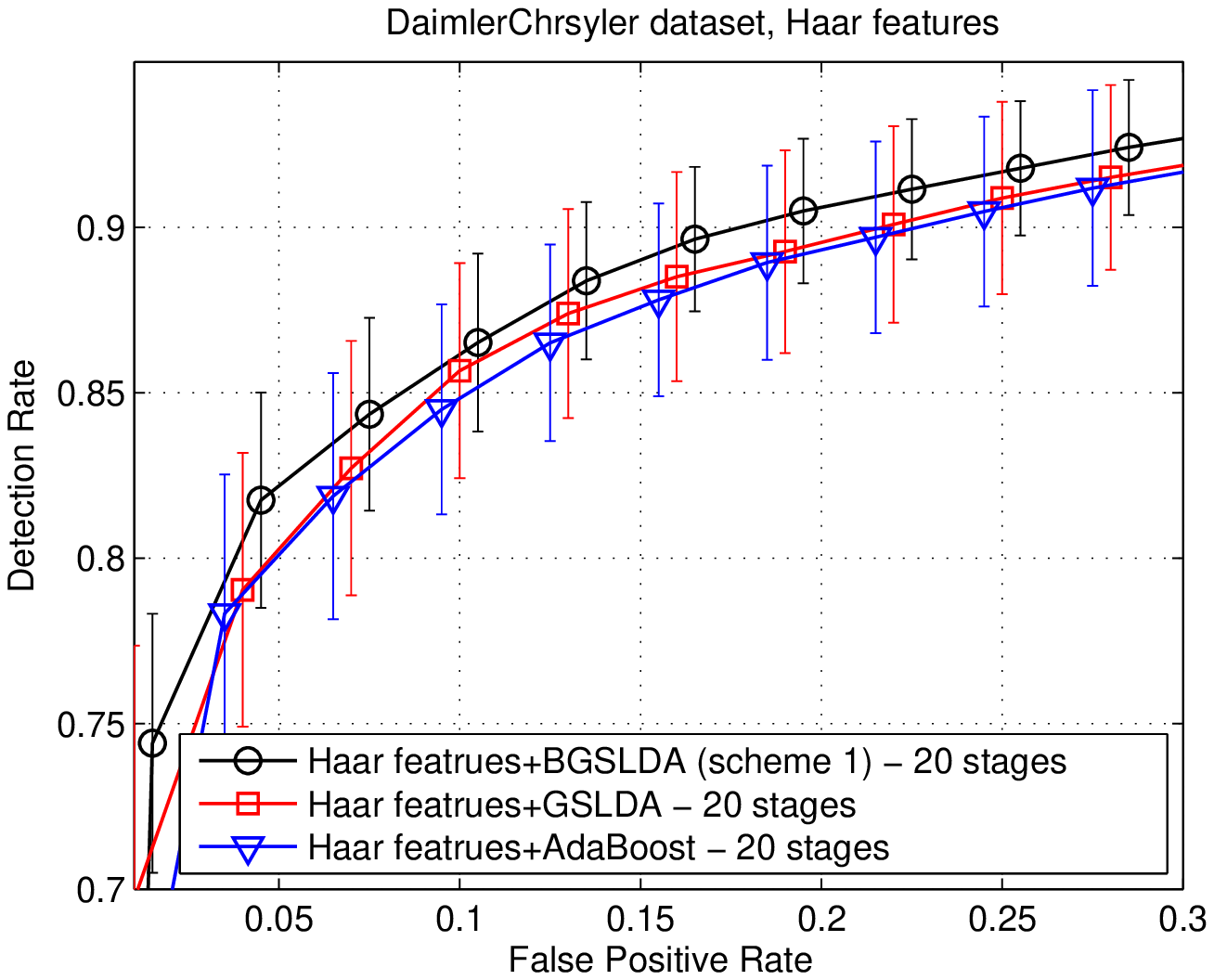}
            \caption{Pedestrian detection performance comparison on the Daimler-Chrysler
            pedestrian dataset \cite{Munder2006Pedestrian}.}
            \label{fig:Daimler1}
        \end{figure}

            In this experiment, we evaluate the performance of our techniques on Daimler-Chrsyler
            pedestrian dataset \cite{Munder2006Pedestrian}.  The dataset contains a set of extracted
            pedestrian and non-pedestrian samples which are scaled to size $18 \times 36$ pixels.
            The dataset consists of three training sets and two test sets. Each training set
            contains  $4,800$ pedestrian examples and $5,000$ non-pedestrian examples. Performance
            on the test sets is analyzed similarly to the techniques described in
            \cite{Munder2006Pedestrian}. For each experiment, three different classifiers are
            generated. Testing all three classifiers on two test sets yields six different ROC
            curves. A $95\%$ confidence interval of the true mean detection rate is given by the
            t-distribution. We conducted three experiments using Haar-like features trained with
            three different classifiers:
            AdaBoost, GSLDA and BGSLDA (scheme $1$). The experimental
            setup is similar to the previous experiments.

            Fig.~\ref{fig:Daimler1} shows detection results of different classifiers.
            Again, the ROC curves show that
            LDA classifier outperforms AdaBoost classifier at all false positive rates.
            Clearly these curves are consistent with those on face datasets.

            \subsubsection{Pedestrian Detection on the INRIA dataset with Covariance Features}

        We also conduct experiments on INRIA pedestrian datasets. We compare the performance of our method with
        other state-of-the-art results. 
        The INRIA dataset \cite{Dalal2005HOG} consists of one training set and one test
set. The training set contains $2,416$ mirrored pedestrian examples and $1,200$ non-pedestrian
images. The pedestrian samples were obtained from manually labeling images taken 
at various time of the days and various locations. The pedestrian samples are mostly in
standing position. A border of $8$ pixels is added to the sample in order to preserve contour
information. All samples are scaled to size $64 \times 128$ pixels. The test set contains $1,176$
mirrored pedestrian examples extracted from $288$ images and $453$ non-pedestrian test images.

Since, Haar-like features perform poorly on this dataset, we apply covariance features instead of
Haar-like features \cite{Tuzel2007Human,Paisitkriangkrai2008Fast}. However, decision stump can not be directly
applied since the algorithm is not applicable to multi-dimensional data. To overcome this problem,
we apply LDA that projects a multi-dimensional data onto a $1$D
space first. In brief, we stack covariance features and project them onto $1$D space. Decision
stumps are then applied as weak classifiers. Our training technique is different from \cite{Paisitkriangkrai2008Fast}. 
\cite{Paisitkriangkrai2008Fast} applied AdaBoost with weighted linear discriminant analysis
(WLDA) as weak classifiers. The major drawback of \cite{Paisitkriangkrai2008Fast} is a slow training time. Since,
each training sample is assigned a weight, weak classifiers (WLDA) need to be trained $T$ times,
where $T$ is the number of boosting iterations. In this experiment, we only train weak classifiers
(LDA) once and store their projected result into a table. Because most of the training time in
\cite{Paisitkriangkrai2008Fast} is used to train WLDA, the new technique requires only $\frac{1}{T}$ training
time as that of \cite{Paisitkriangkrai2008Fast}. After we project the multi-dimensional covariance features onto
a $1$D space using LDA, we train decision stumps on these $1$D features. In other words, we replace
line $4$ and $5$ in Algorithm~\ref{ALG:GSLDA} with Algorithm~\ref{ALG:GSLDA_TRAIN_MULTI}.

\SetVline
\linesnumbered
\begin{algorithm}[t]
\caption{The algorithm for training multi-dimensional features.}
\begin{algorithmic}
\small{

   \ForEach{multi-dimensional feature}
   {
     1. Calculate the projection vector with LDA and project the 
         multi-dimensional feature to $1$D space;
     \\
     2. Train decision stump classifiers to find an optimal 
         threshold $ \theta $ using positive and negative training set;
   }
   
} 
\end{algorithmic}
\label{ALG:GSLDA_TRAIN_MULTI}
\end{algorithm}

        In this experiment, we generate a set of over-complete rectangular covariance filters and
        subsample the over-complete set in order to keep a manageable set for the training phase.
        The set contains approximately $45,675$ covariance filters. In each stage, weak classifiers
        are added until the predefined objective is met. We set the minimum detection rate to be
        $99.5\%$ and the maximum false positive rate to be $35\%$ in each stage. The cascade
        threshold value is then adjusted such that the cascade rejects $50\%$ negative samples on
        the training sets. Each stage is trained with $2,416$ pedestrian samples and $2,500$
        non-pedestrian samples. The negative samples used in each stage of the cascades are
        collected from false positives of the previous stages of the cascades. 

\begin{figure}[t!]
	  \centering
		\includegraphics[width=0.42\textwidth,clip]{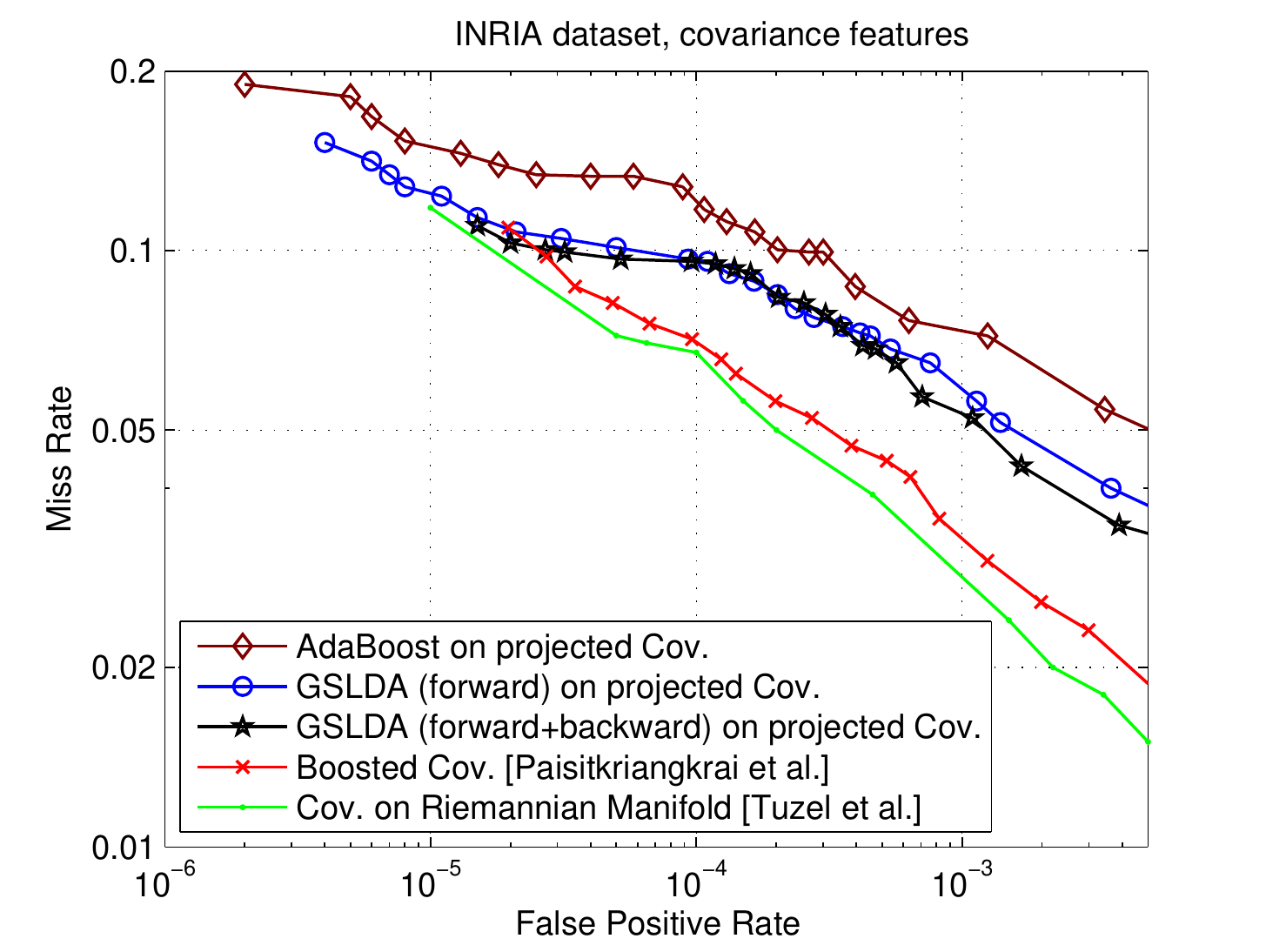}
		\caption{
		         Pedestrian detection performance comparison of $1$D covariance features (projected
                 covariance) 
		         trained using AdaBoost and GSLDA on the INRIA dataset \cite{Dalal2005HOG}.
			    }
	 \label{fig:INRIA}
  \end{figure}

        Fig.~\ref{fig:INRIA} shows a comparison of our experimental results on learning $1$D
        covariance features using AdaBoost and GSLDA. The ROC curve is generated by adding one
        cascade level at a time.  From the curve, GSLDA classifier outperforms AdaBoost classifiers
        at all false positive rates. The results seem to be consistent with our results reported
        earlier on face detection. On a closer observation, our simplified technique performs very
        similar to existing covariance techniques 
        \cite{Tuzel2007Human,Paisitkriangkrai2008Fast} at low false
        positive rates (lower than $10^{-5}$). This method, however, seems to perform poorly at high false
        positive rates. Nonetheless, most real-world applications often focus on low false
        detections. Compared to boosted covariance features, the training time of cascade
        classifiers is reduced from {\em weeks to days} on a standard PC.

  \section{Conclusion}
\label{sec:conclusion}

        In this work, we have proposed an alternative approach in the
        context of visual object detection. 
        The core of the new framework is greedy sparse linear
        discriminant analysis (GSLDA)
        \cite{Moghaddam2007Fast}, which aims to maximize the 
        class-separation criterion.
        On various datasets for face detection and pedestrian detection,
        we have shown that this technique
        outperforms AdaBoost when the distribution of positive and
        negative samples is highly skewed. 
        To further improve the
        detection result, we have proposed a boosted version GSLDA, 
        which combines boosting re-weighting scheme with decision
        stumps used for training the GSLDA algorithm. 
        Our extensive experimental results show that the
        performance of BGSLDA is better than that of AdaBoost at
        a similar computation cost.

        Future work will focus on the search for more efficient weak
        classifiers and on-line updating the learned model.


\bibliographystyle{ieee}

\bibliography{journal,conf,book,misc}

\end{document}